\documentclass[10pt]{IEEEtran}

\usepackage{cite}
\usepackage{epsfig}
\usepackage{url}
\usepackage{subfigure}

\begin{document}

\markboth{
}{
  Tan and Clarke: 
  A Family of Rank Similarity Measures based on
  Maximized Effectiveness Difference
}

\title{
  A Family of Rank Similarity Measures based on\\
  Maximized Effectiveness Difference
}

\author{
  Luchen Tan and Charles L. A. Clarke
  \thanks{
    The authors are with the
    School of Computer Science, University of Waterloo, Canada
  }
}

\maketitle

\begin{abstract}
Rank similarity measures provide a method for quantifying differences
between search engine results without the need for relevance judgments.
For example, the providers of a search service might use such measures
to estimate the impact of a proposed algorithmic change across a large
number of queries~---~perhaps millions~---~identifying those queries where
the impact is greatest.
In this paper, we propose and validate a family of rank similarity measures,
each derived from an associated effectiveness measure.
Each member of the family is based on the maximization of effectiveness
difference under this associated measure.
Computing this maximized effectiveness difference (MED) requires the solution
of an optimization problem that varies in difficulty,
depending on the associated measure.
We present solutions for several standard effectiveness measures,
including nDCG, MAP, and ERR.
Through an experimental validation, we show that MED reveals meaningful
differences between retrieval runs.
Mathematically, MED is a metric, regardless of the associated measure.
Prior work has established a number of other desiderata for rank similarity
in the context of search, and we demonstrate that MED satisfies these
requirements.
Unlike previous proposals, MED allows us to directly translate assumptions
about user behavior from any established effectiveness measure to create
a corresponding rank similarity measure.
In addition, MED cleanly accommodates partial relevance judgments,
and if complete relevance information is available, it reduces to a
simple difference between effectiveness values.
\end{abstract}

\begin{keywords}
  Search, rank similarity, information retrieval, effectiveness measures,
  search engines
\end{keywords}

\section{Introduction}

\PARstart{E}{ven}
minor modifications to ranking algorithms, training data, or document
collections may cause unpredictable changes to the quality of search engine
results.
For any given query, documents may move up and down the ranked list.
New documents may appear; others may disappear.

Traditionally, changes in search engine results are quantified by way of
retrieval effectiveness measures,
such as normalized discounted cumulative gain (nDCG),
mean average precision (MAP),
and expected reciprocal rank (ERR)~\cite{kk02,cmzg09}.
These measures are computed over ranked lists before and after a modification.
The difference in their values~---~typically averaged over a number of
queries~---~indicates the magnitude of the change.

Unfortunately, the computation of these measures depends on the existence of
explicit relevance judgments.
For a given query, we must know whether or not each ranked document is
relevant to the query, and for some measures we must also know the degree
to which each document is relevant to the query
(e.g., definitive, excellent, good, fair, bad, or detrimental~\cite{najork07}).
These judgments may be created by direct human assessment of relevance or
inferred indirectly from clickthrough logs and other user interaction
data~\cite{cz09,dl10,ehvs12,hzcwy11,jgphg05,sj10}.
Creation of these judgments involves either substantial effort on the part of
assessors or large volumes of interaction data,
limiting the number of queries over which the measures may be computed.
Moreover, when modifications surface previously unseen and unjudged documents,
these documents must be assessed immediately,
or the accuracy of the effectiveness measures may be compromised.

Alternatively, changes in search results may be quantified by way of rank
correlation coefficients and other rank similarity measures,
avoiding the need for explicit relevance judgments.
For example, the providers of a search service might use such measures
to estimate the impact of a proposed change across large numbers of
queries~---~perhaps millions of them~---~identifying those
where the potential impact is greatest.
While standard rank correlation coefficients~---~such as Spearman's $\rho$
or Kendall's $\tau$~---~might be applied for this purpose,
several researchers have proposed specialized rank similarity measures that
better reflect the requirements of search.

Some of this prior work focuses primarily on the problem of comparing search
result lists~\cite{chris04,slc10,mel09,kv10},
a problem which we directly address in this paper.
Most notably, Webber et al.~\cite{wmz10} carefully analyze the requirements
for comparing ranked lists across a wide range of applications,
including search results.
Other work focuses primarily on the distinct, but related, problem of
comparing system rankings under different effectiveness
measures~\cite{ben09,yar08},
a problem which we do not directly address in this paper.
We note, however, that progress on one problem may translate into progress
on the other.

Webber et al.\ identify three key characteristics 
that particularly apply to comparisons between search results.
First, users tend to focus on the top-ranked results, and more rarely view
results deep in the list~\cite{jr06,jgphg05}.
Thus, a rank similarity measure for search results must be \emph{top-weighted},
placing greater emphasis on early results, and lesser emphasis on later results.
A change in the third result should have greater impact than a change in the
103rd.
Second, for a given query, search engines do not typically rank more than a
tiny fraction of the collection.
Thus, a similarity measure for search results must handle \emph{incomplete}
rankings, where a document appearing in one list may not appear in the other.

Finally, since users may stop at any point in a ranked list, either because
their information need is satisfied or because their patience is exhausted,
a rank similarity measure for search results must appropriately handle
\emph{indefinite} rankings.
According to Webber et al.\ the measure ``should not arbitrarily assign a
cutoff depth, but be consistent for whatever depth is available.''
Suppose the ranked lists are truncated at an arbitrary depth $K$ and we
compute the measure at that depth.
The value of the measure at $K$ should allow us to make predictions about the
value of the measure if it were computed at depths greater than $K$,
allowing us to compute bounds on the amount it can change as the depth of
computation is increased.
For example, perhaps the measure would only increase, or stay the same, as the
depth increased (i.e., adding more documents could not make the lists less
similar).

To satisfy the requirements implied by these three characteristics,
Webber et al.\ proposed a new rank similarity measure, which they call
\emph{rank biased overlap} (RBO).
Given two ranked lists, $A$ and $B$,
let $A_{1:k}$ denote the top $k$ documents in $A$,
and let $B_{1:k}$ denote the top $k$ documents in $B$.
Define the {\em overlap} between $A$ and $B$ at depth $k$ as the
size of the intersection between these lists at depth $k$
(i.e., $\left|A_{1:k} \cap B_{1:k}\right|$)
and define the {\em agreement} between $A$ and $B$ at depth $k$ as
the overlap divided by the depth.
Webber et al.\ define RBO as a weighted average of agreement across depths,
where the weights decay geometrically with depth, reflecting the
requirement for top weighting:
\begin{equation}
\label{RBO}
\mbox{RBO} =
  (1 - \psi)
    \sum_{k=1}^{\infty}
      \psi^{k - 1} \frac{\left|A_{1:k} \cap B_{1:k}\right|}{k}.
\end{equation}
In this equation, the parameter $0 < \psi < 1$ represents user persistence,
with larger values representing a more patient user.
When computing RBO for the comparative experiments reported in this paper,
we set $\psi = 0.9$, a typical choice.
The normalization factor $(1 - \psi)$ serves to map the value of RBO into the
range $[0:1]$.
In practice, RBO is computed down to some fixed depth $K$,
which replaces $\infty$ in the equation, reflecting the indefinite and
incomplete nature of the ranked lists.

Webber et al.\ surveyed a large number of rank similarity measures, and
argued that only RBO fully satisfies the requirements of ranked search
results.
In creating RBO, Webber et al.\ drew inspiration from the user model
incorporated into the rank-bias precision (RBP) effectiveness measure
proposed by Moffat and Zobel~\cite{mz08}.
This model imagines a user scanning a ranked list, starting at the top.
After scanning a result, the user proceeds to the next result with probability
$\psi$ and stops with probability $1 - \psi$.
This simple user model provides the basis for the weights appearing in both
RBP and RBO.

This close connection with the RBP effectiveness measure helps to justify the
application of RBO to search results.
In general, given the enormous effort made by the search community to develop
and validate effectiveness measures, it seems reasonable to exploit this
existing work to guide the creation of rank similarity measures targeted at
search results.
Taking this observation a step further, we propose a family of distance
measures, each directly derivable from an associated effectiveness measure.
The core idea is straightforward~\cite{ca05,cas06,ben08}:
\begin{quote}
Given two ranked lists, $A$ and $B$, what is the maximum difference in
their effectiveness scores possible under a specified effectiveness measure?
\end{quote}
We call this family of distance measures
\emph{Maximized Effectiveness Difference} (MED).
Since we typically normalize MED into the range $[0:1]$, the value $1 - MED$
provides an corresponding family of rank similarity measures.

Like RBO, MED is a metric in the strict mathematical sense of defining a
distance between two ranked lists (see Section~\ref{SNotation}).
The value of this distance varies with the associated effectiveness measure.
Unlike most similarity measures, MED is not a dimensionless quantity,
since its value can be interpreted in terms of the associated effectiveness
measure.

Like RBO, MED transfers understanding and assumptions about user behavior from
an existing effectiveness measure to measure rank similarity.
Unlike RBO, MED provides a method for transforming any effectiveness
measure into a rank similarity measure.
Both MED and RBO derive their approach to top weighting from their associated
effectiveness measure, along with an ability to handle incomplete rankings,
satisfying the requirements of search.
MED is monotonically decreasing with increasing depth of evaluation,
satisfying the requirements for indefinite rankings.
In addition, MED can take appropriate advantage of known relevance information
(see Section~\ref{SNotation}).
It is straightforward to show, from its definition, that introducing known
relevance information always decreases the MED distance between two ranked
lists.
Moreover, if complete relevance information is known, MED reduces to a
difference in effectiveness scores.

For an effectiveness measure normalized to the range $[0:1]$,
the value of MED will also fall into the range $[0:1]$.
As we discuss later in the paper,
for some effectiveness measures this normalization requires knowledge of
collection parameters, e.g., the total number of relevant documents.
Since such information cannot be known without complete relevance information,
we choose values for these parameters that normalize MED into the
range $[0:1]$.

To compute MED, we must determine an assignment of relevance values to the
documents appearing in the two lists that maximizes the difference in their
scores under the specified effectiveness measure, an optimization process
that will vary with the effectiveness measure.
For some measures, such as nDCG and RBP, the computation of MED is
straightforward (see Section~\ref{SDotMED}).
For other measures, including MAP and ERR, the computation of MED requires
the solution of more difficult optimization problems
(see Sections~\ref{SMAPMED} and~\ref{SERRMED}).
Either way, MED reflects meaningful differences between ranked lists
(see Section~\ref{SValidation})
and provides substantial advantages over other rank similarity measures
on search results (see Section~\ref{SComparison}).
Finally, we consider how to extend MED beyond ranked lists,
to more other effectiveness measures~(see Section~\ref{SBeyond}).

\section{Basic Notation and Properties}
\label{SNotation}

We compute MED by maximizing the effectiveness difference between two ranked
lists, $A$ and $B$.
If $S(A)$ is the score for list $A$ under some effectiveness measure,
and $S(B)$ is the score of list $B$ under the same measure, then we seek to
assign relevance values to the documents in $A$ and $B$ to maximize
\begin{equation}
\mbox{MED}(A,B) = \left|S(A) - S(B)\right|.
\end{equation}

In the remainder of the paper, we assume that scores produced by effectiveness
measures are always non-negative (and we are not aware of any
established effectiveness measure that can produce negative values).
Thus, without loss of generality, we present algorithms that maximize
\begin{equation}
\label{EEE}
  S(A) - S(B).
\end{equation}
The overall maximum can be computed by swapping the lists and reapplying the
algorithm.

For optimization purposes, we represent $A$ as a vector of variables
\begin{equation}
  A = \langle a_1, a_2, ..., a_K \rangle,
\end{equation}
where $a_i$,$1 \leq i \leq K$, represents the relevance value assigned to the
document at rank $i$ in list $A$. 
Similarly, we represent $B$ as a vector of variables
\begin{equation}
  B = \langle b_1, b_2, ..., b_K\rangle,
\end{equation}
where $b_j$, $1 \leq j \leq K$, represents the relevance value assigned to the
document at rank $j$ in list $B$. 
Since a document may appear in both lists, we also have a set of constraints
in the form
\begin{equation}
  a_n \equiv b_m,
\end{equation}
indicating that the same document appears at rank $n$ in list $A$ and at rank
$m$ in list $B$, so that the same relevance value must be assigned to both
variables.
Since a document will appear at most once in each list, a given variable will
appear in at most one constraint.

We refer to a variable that appears in a constraint as a \emph{bound variable}.
If a variable does not appear in any constraint~---~corresponding to a document
that appears in only one list~---~we refer to it as a \emph{free variable}.
In the case that relevance information for a document is known from existing
judgments, we assign this known value to its corresponding variable(s),
which remain unchanged during maximization.
We refer to these variables as \emph{predetermined variables}.

For all effectiveness measures we consider in this paper, a relevance value
is a number in the range $[0:1]$.
For some measures~---~such as MAP~---~relevance is a binary property
of a document, and relevance values are either $0$ or $1$.
For graded relevance measures~---~such as nDCG and ERR~---~relevance can
be one of several values from $0$ up to a maximum grade $r_G$ (i.e., the
possible relevance grades are $0 = r_0 < r_1 < r_2 < ... < r_G \leq 1$).
These grades indicate the level to which a document is judged
relevant to the query (e.g., definitive, excellent, good, fair, etc.).
For simplicity, we treat binary relevance as a special case of graded
relevance, with two grades: $r_0 = 0$ and $r_G = r_1 = 1$.
To aid understanding, a relevance value may be interpreted as the
probability that a user viewing the corresponding document will consider it
to be relevant~\cite{cztr08,cmzg09,ccsa11}, but this interpretation is not
explicitly required in this paper.

Mathematically, MED is a metric, regardless of the associated
effectiveness measure.
Non-negativity, identity and symmetry are straightforward.
To demonstrate the triangle inequality, consider three ranked lists
$A$, $B$, and $C$.
Let $A'$ and $B'$ represent the assignment of relevance values that
maximizes the effectiveness difference between $A$ and $B$,
i.e., $\mbox{MED}(A,B) = \left|S(A') - S(B')\right|$.
Let $C'$ be any assignment of relevance values to $C$ that is consistent
with $A'$ and $B'$, such that documents are assigned the same relevance values
in all three lists.
Let $A''$ and $C''$ represent the assignment of relevance values that
maximizes the effectiveness difference between $A$ and $C$.
Let $B'''$ and $C'''$ represent the assignment of relevance values that
maximizes the effectiveness difference between $B$ and $C$.
Now,
\begin{eqnarray}
\mbox{MED}(A,B)
  &=& \left|S(A') - S(B')\right|                                       \\ 
  &=&   \left|S(A') - S(C') + S(C') - S(B')\right|                     \nonumber \\
  &\leq& \left|S(A') - S(C')\right| + \left|S(C') - S(B')\right|       \nonumber \\
  &\leq& \left|S(A'') - S(C'')\right| + \left|S(C''') - S(B''')\right| \nonumber \\
  &=&   \mbox{MED}(A,C) + \mbox{MED}(C,B).                             \nonumber
\end{eqnarray}
The second-last step holds by the definition of MED, as maximizing effectiveness
difference.

The degree to which MED satisfies our requirements with respect to
top-weighting, indefinite rankings, and incomplete rankings,
depends upon the associated effectiveness measure.
While we are not aware of any established effectiveness measure that
cannot handle incomplete rankings,
a few measures~---~such as precision@$k$~---are not top weighted.
Moreover, several effectiveness measures~---~such as precision@$k$
and nDCG@k~---~are parameterized by a depth $k$.
Changing the value of $k$ effectively creates a new measure.
For such effectiveness measures, extending the ranked lists beyond $k$
does not change the value of the measure, or the value of MED.
For these measures, MED appropriately supports indefinite rankings only
to depths less than or equal to $k$.
When $K < k$, we compute MED by filling ranks $K + 1$ to $k$ with free
variables in both lists.
Other measures~---such as RBP and ERR~---~are not parameterized by depth
and are notionally computed to infinity,
providing stronger support for indefinite rankings.

\section{Simple Dot Product Measures}
\label{SDotMED}

In this section, we maximize Equation~\ref{EEE} for a class of simple but
widely used effectiveness measures.
These measures may be expressed as a normalized dot product between a
vector of relevance values and a vector of rank-based discount values.
The class includes RBP and nDCG, along with other measures~\cite{mst12}.

Let $C$ be a ranked list (which could be either $A$ or $B$) represented by
a vector $C = \langle c_1, c_2, ... \rangle$ of relevance values.
Let $D = \langle d_1, d_2, ... \rangle$ be a vector of discount values,
where the value of $d_i$ depends only on the rank, $i$.
Effectiveness is computed as the normalized dot product of $C$ and $D$:
\begin{equation}
\label{EDot}
S(C) = \frac{C \cdot D}{{\cal N}},
\end{equation}
where the normalization factor ${\cal N}$ is a constant, which may depend
on assumptions about the user or on characteristics of the document collection,
such as the number of relevant documents it contains at each relevance grade.
Generally, normalization serves to map the value of the measure into the
range $[0:1]$.
To compute MED, we require ${\cal N} > 0$ (and we are not aware of any
established effectiveness measure where this requirement does not hold).

Each $d_i$ in the discount vector may be interpreted as the probability that
a user scanning the search results will reach rank $i$, and hence view the
document at that rank~\cite{cmzg09,ccsa11,mz08},
but this interpretation is not explicitly required in this paper.
To compute MED, we require discount values to be non-negative and to decrease
monotonically with increasing rank, i.e., $d_i \geq d_j$ if $i < j$ (and
we are not aware of any established effectiveness measures where this
requirement does not hold).

Maximizing Equation~\ref{EEE} for simple dot product measures is
straightforward.
First, we set all predetermined variables in both lists to their known values.
Second, we set all free variables in $A$ to $r_G$ and all free variables
in $B$ to $r_0 = 0$.
If $S(A) - S(B)$ is maximized, then all free variables in $A$ must have value
$r_G$, for otherwise we could increase $S(A) - S(B)$ by increasing the
value of these free variables.
Similarly, if $S(A) - S(B)$ is maximized, then all free variables in $B$
must have value $0$.

Finally, given a constraint $a_n \equiv b_m$, the contribution of these
variables to $S(A) - S(B)$ is
\begin{equation}
\frac{a_n d_n - b_m d_m}{{\cal N}}.
\end{equation}
Since discount decreases monotonically with increasing rank,
$S(A) - S(B)$ is maximized by setting
$a_n = b_m = r_G$ if $n < m$, or $a_n = b_m = 0$ if $n > m$.
If $n = m$ then these variables contribute nothing to the value of
$S(A) - S(B)$ and can be set to any value.

For example, we may express precision@$k$ as a simple dot product measure
using a vector of binary relevance values, a normalization factor of $k$,
and a discount with the first $k$ values set to one and the remaining
values set to $0$: 
\begin{equation}
  D =
    \langle
      d_1 = 1, d_2 = 1, ...,d_k = 1, d_{k+1} = 0, ...
    \rangle.
\end{equation}
In the case that the ranked lists are not fully specified to depth $k$, we
complete them with arbitrarily chosen free variables.
Applying the relevance assignment process above, we see that MED for
the precision@$k$ effectiveness measure (which we call MED-precision@$k$)
is just one minus the overlap between $A$ and $B$ at depth $k$:
\begin{equation}
  1 - \frac{\left|A_{1:k} \cap B_{1:k}\right|}{k}.
\end{equation}
If there are no predetermined variables, then for simple dot measures
it is not hard to show that maximizing $S(B) - S(A)$ produces the same value
as maximizing $S(A) - S(B)$.
In this case, there is no need to swap lists and recompute.

\subsection{Computing MED-RBP}

Moffat and Zobel~\cite{mz08} define the formula for computing rank biased
precision (RBP) as:
\begin{equation}
S(C) = (1 - \psi) \sum_{i=1}^\infty c_i \psi^{i - 1},
\end{equation}
where $C = \langle c_1, c_2, ... \rangle$ is a vector of graded relevance
values, although the definition works equally well for binary values.
For the experiments reported in this paper, we assume $r_G = 1$.
As it does in RBO, the parameter $0 < \psi < 1$ represents user persistence,
with larger values representing a more patient user.
We may easily express this formula in the form of Equation~\ref{EDot},
making it straightforward to compute MED for the RBP measure (MED-RBP).
When computing MED-RBP for the experiments reported in this paper,
we set $\psi = 0.9$, a typical choice.

MED-RBP provides strong properties in support of indefinite rankings.
For ranked lists specified only to depth $K$, we compute MED-RBP
down to infinite depth by assuming arbitrary free variables in both lists
below $K$.
To maximize $S(A)-S(B)$, in list $A$ we set these free variables to $1$,
and in list $B$ we set these free variable to $0$, so that we maximize:
\begin{eqnarray}
  S(A) - S(B)
    &=&
      (1 - \psi) \left(
	\sum_{i=1}^K (a_i - b_i) \psi^{i - 1} +
	\sum_{i=K+1}^\infty \psi^{i - 1}
      \right) \nonumber \\
    &=&
      (1 - \psi) \left(\sum_{i=1}^K (a_i - b_i) \psi^{i - 1}\right) + \psi^K.
\end{eqnarray}
If the ranked lists are later specified to greater depth,
increasing $K$ and potentially introducing new bound and predetermined
variables, MED cannot increase.
Moreover, the MED-RBP distance cannot decrease by more than $2\psi^K$ as the
depth goes to infinity.

\subsection{Computing MED-nDCG}

At the time J\"{a}rvelin and Kek\"{a}l\"{a}inen~\cite{kk02} created
normalized discounted cumulative gain (nDCG) no other established
effectiveness measure accommodated graded relevance values.
Since then, nDCG has become widely used for Web-related research.
In this paper, we work with a version of nDCG that has become standard
in the research literature and through industry practice~\cite{cmzg09,najork07}:
\begin{equation}
\label{EnDCG}
  S(C) =
    \left(\frac{1}{\mbox{\em ideal DCG}} \right)
      \sum_{i = 1}^k \frac{c_i}{\log \left(i + 1\right)}.
\end{equation}
$C = \langle c_1, c_2, ... \rangle$ is a vector of relevance
values, where each $c_i$ is one of $r_0$ ... $r_G$.
When computing nDCG for our experiments, we set $k = 20$, a typical choice.

For nDCG@k, relevance values are computed using the formula~\cite{cmzg09}:
\begin{equation}
\label{Erel}
  r_j = \frac{2^j - 1}{2^G}, \ \ j = 0,...,G.
\end{equation}
For our experiments, the test collection employs three relevance grades,
so that $r_0 = 0$, $r_1 = 1/4$, and $r_2 = r_G = 3/4$.
These grades indicate documents that are judged to be non-relevant, relevant,
and highly relevant, respectively.

We may express Equation~\ref{EnDCG} in the form of Equation~\ref{EDot},
with a discount vector of
\begin{equation}
  D = \langle 1,..., 1/{\log \left(i + 1\right)},... \rangle.
\end{equation}
J\"{a}rvelin and Kek\"{a}l\"{a}inen call the dot product $C \cdot D$
\emph{discounted cumulative gain} (DCG).
DCG is then normalized by \emph{ideal DCG} to give nDCG.

Ideal DCG is a constant, defined as the maximum DCG achievable over the
collection, which can be determined by ranking all the most relevant
documents first, followed by the next most relevant, and so on.
In theory, determining ideal DCG may require exhaustive judging, which
is rarely feasible on realistically sized collections.
In practice, ideal DCG is usually estimated from the known relevant documents
surfaced during an retrieval experiment.
If changes to ranking algorithms surface new relevant documents,
the estimate of ideal DCG may grow and the value of nDCG may drop,
even when the changes improve the algorithm.
When applying nDCG as an effectiveness measure, care must be taken to account
for this potential growth in ideal DCG.

When computing MED for the nDCG measure (MED-nDCG), we may have no judgments
at all.
Fortunately this problem can be avoided to some extent.
The definition of MED implicitly assumes that the collection may contain an
arbitrary number of maximumly relevant documents.
Following this assumption, we define
\begin{equation}
  \mbox{\em ideal nDCG}
    = {\cal N}
    = \sum_{i = 1}^k \frac{r_G}{\log \left(i + 1\right)}.
\end{equation}
Using this normalization factor, the value of MED-nDCG will fall in the
range $[0:1]$.

Admittedly, we do lose one property of MED with this approach.
The calculation of nDCG requires the estimation of a constant that depends
solely on the collection, not on the ranked lists being measured.
With no relevance judgments available,
we have no estimate for the value of this constant.
By normalizing it away, as we do above,
we lose the property that MED can be interpreted as an actual maximum
difference in effectiveness values.
MED-nDCG continues to be a metric in the mathematical sense,
able to measure distances between rankings,
but these distances are scaled by an unknown constant.
In this regard, MED-nDCG is no different than existing correlation coefficients,
including RBO, which are also dimensionless quantities.

Alternatively, for some queries we may know, or be able to reasonably guess,
properties of the collection that would impact normalization.
For example, we may known through the application of a query type classifier
that a given query is navigational, and we may know from experience that a
navigational query typically has one highly relevant document and a small
number~---~no more than a dozen, say~---~of marginally relevant documents.
Under these assumptions the computation of MED-nDCG would require a different
normalization constant and additional constraints on the optimization,
limiting the values of variables.
We leave the exploration of this idea for future work.

\section{Computing MED-MAP}
\label{SMAPMED}

Mean average precision (MAP) is defined over a vector
$C = \langle c_1, c_2, ... \rangle$ of binary relevance values as:
\begin{equation}
  \label{EMAP}
  S(C)
    \ =\  \frac{1}{R} \sum_{i = 1}^k (c_i \cdot \mbox{\ precision@$i$})
    \ =\  \frac{1}{R} \sum_{i = 1}^k \frac{c_i}{i} \sum_{j = 1}^i c_j,
\end{equation}
where $R$ indicates the number of relevant documents in the collection,
and $k$ an arbitrary maximum depth for computation.
Although it is rarely made explicit in the research literature as MAP@$k$,
this maximum depth is as important for MAP as it is for precision@$k$ and
nDCG@$k$.
Changing $k$ effectively creates a different measure.
When computing MAP for our experiments, we set $k = 100$, a typical choice.

Like ideal nDCG, determining $R$ theoretically requires exhaustive judging.
However, as we did for ideal nDCG, when computing MED for the MAP measure
(MED-MAP) we assume that the collection contains an arbitrary number of
relevant documents.
To compute MED-MAP, we replace $R$ with $k$ in Equation~\ref{EMAP}, which
guarantees that the value of MED-MAP will fall in the range $[0:1]$:
\begin{equation}
  \label{EMAP2}
  S(C) = \frac{1}{k} \sum_{i = 1}^k \frac{c_i}{i} \sum_{j = 1}^i c_j.
\end{equation}

Unfortunately, Equation~\ref{EMAP2} is quadratic in its relevance values and
does not fit the simple dot product form assumed in Section~\ref{SDotMED}.
Maximizing $S(A) - S(B)$ requires a little more effort than it does
for those measures.
Fortunately, this maximization problem can be re-expressed as a quadratic
0-1 optimization problem, a heavily researched and well understood
class of problems~\cite{rrw09,bea98}.
Our goal is to assign relevance values to the documents in lists $A$ and $B$
that maximizes
\begin{equation}
\label{EMAP3}
S(A) - S(B) = \frac{1}{k} \left(
  \sum_{i = 1}^k \frac{a_i}{i} \sum_{j = 1}^i a_j - 
  \sum_{i = 1}^k \frac{b_i}{i} \sum_{j = 1}^i b_j
\right).
\end{equation}
If $S(A) - S(B)$ is maximized, it is not hard to show that the free variables
in $A$ must be set to one, and the free variables in $B$ must be set to zero.
Of course, predetermined variables must be set to their known values.

After setting the values for free and predetermined variables,
our next step is to replace each pair of variables appearing in a
constraint with a single variable.
To do this, we create a combined variable vector
$Z = \langle z_1, z_2, ..., z_{k'} \rangle$,
where $k' \leq k$ is the number of constraints.
Each document appearing in both in $A$ and $B$ corresponds to a single
variable in $Z$.
The ordering of $Z$ is arbitrary.
Equation~\ref{EMAP3} can now be re-written in the form
\begin{equation}
\label{Emaxcut}
  Z^T Q Z + L^T Z + F,
\end{equation}
where $Q$ is a matrix of order $k'$, L is a vector of dimension $k'$,
and $F$ is a constant.
Equation~\ref{Emaxcut} is the standard form for quadratic 0-1 optimization,
a heavily studied NP-complete problem.
Quadratic 0-1 optimization is equivalent to the weighted max-cut problem,
one of Karp's original 21 NP-complete problems.

To approximate MED-MAP, we implemented a version of
\emph{tabu search}~\cite{bea98},
a standard local search method that creates and maintains a set of
disallowed (or \emph{tabu}) moves to avoid repeated visits to suboptimal
solutions.
Unfortunately~---~while the computation of MED for dot product measures is
essentially instantaneous~---~the computation of MED-MAP may require a
second or so on a typical desktop machine.

\section{Computing MED-ERR}
\label{SERRMED}

Expected reciprocal rank (ERR) is based on the \emph{cascade model} of
user browsing behavior over search results~\cite{cmzg09,cztr08}.
The model implicitly assumes that the user is seeking a single relevant
document.
After entering a search query and receiving a result list,
the user scans the list, starting at the first result.
With probability $c_1$ the user finds the information she seeks and stops
browsing.
Otherwise, with probability $1 - c_1$, she continues to the second result,
and so on.
In general, if the user reaches the result at rank $i$, she finds the
information she seeks with probability $c_i$, or continues on to the result
at rank $i+1$ with probability $1 - c_i$.
Thus, the probability that the user reaches rank $i$ is
\begin{equation}
\label{Ereach}
  \prod_{j = 1}^{i - 1} (1 - c_j).
\end{equation}
ERR is then defined as the expected reciprocal rank where the user's
information need is satisfied:
\begin{equation}
  \label{EERR}
  S(C) = \sum_{i = 1}^\infty \frac{c_i}{i} \prod_{j = 1}^{i - 1} (1 - c_j).
\end{equation}
ERR uses the same relevance grades as nDCG, as defined by Equation~\ref{Erel}.
ERR is not normalized; its value naturally falls into the range $[0:1]$
(with a maximum value $<$ 1.0).

The cascade model is closely related to the user model incorporated into
RBO and RBP, as described in the introduction.
For RBO and RBP, the probability that the user will move from rank $i$ to $i+1$
is constant ($\psi$).
Under the cascade model, this inter-rank transition probability depends on
the relevance of the document at rank $i$, and the probability that the
user will reach rank $i$ depends on the relevance of all the documents
appearing before it.
If a few highly relevant documents appear above rank $i$, the probability
of reaching that rank becomes relatively small.
For example, if $r_G = 3/4$, it takes only four such documents for the
probability in Equation~\ref{Ereach} to drop below 1\%.
After viewing five such documents, less than one in a thousand users will
continue.

Under ERR, if $S(A) - S(B)$ is maximized, it is possible to show that
free variables in $A$ must be set to $r_G$ and free variables in $B$ must be
set to $r_0 = 0$.
Moreover, if $S(A) - S(B)$ is maximized, it is possible to show that
bound variables must either be set to $r_G$ or $0$, apart from cases where
the value of the variable does not change the value of the difference.
In these cases, we can arbitrarily set the variables to $r_G$ or $0$.
We omit proofs for these claims, which are straightforward but tedious.

Thus, in maximizing $S(A) - S(B)$ we set free and bound variables to either
$r_G$ or $0$, allowing us to ignore intermediate relevance grades.
Unfortunately, maximizing $S(A) - S(B)$ still requires us to solve a highly
non-linear optimization problem, with what are effectively 0-1 constraints.
Fortunately, we can take advantage of properties of ERR to efficiently
approximate the solution.
In particular, as we noted previously, the value of ERR largely depends on
the position of the first few relevant documents.
As the user views more and more relevant documents, the probability she
will continue to lower ranks drops exponentially.

Consider the calculation of $S(A)$ for some assignment to the variables in $A$.
We start calculating the summation at rank $i = 1$.
Suppose at rank $k$ we have encountered $p$ variables with value $r_G$.
The sum over the remaining ranks is bounded by
\begin{eqnarray}
    \epsilon
    & = & \sum_{i = k + 1}^{\infty} \frac{c_i}{i} \prod_{j = 1}^{i - 1} (1 - c_j)
    \\
    & \leq & \sum_{i = p + 1}^{\infty} \frac{c_i}{i} \prod_{j = 1}^{i - 1} (1 - c_j)
    \nonumber \\
    & \leq & \sum_{i = p + 1}^{\infty} \frac{r_G}{i} \prod_{j = 1}^{i - 1} (1 - r_G)
    \nonumber \\
    & = & {(1 - r_G)}^p \sum_{i = p + 1}^{\infty} \frac{r_G}{i} \prod_{j = p + 1}^{i - 1} (1 - r_G)
    \nonumber \\
    & < & \frac{r_G {(1 - r_G)}^p}{p + 1} \sum_{i = 0}^{\infty} {(1 - r_G)}^i
    \nonumber \\
    & = & \frac{{(1 - r_G)}^p}{p + 1}.
    \nonumber
\end{eqnarray}
If $p = 5$ and $r_G = 3/4$ then $\epsilon < 0.0002$.
This bound on $S(A)$ is also a bound on $S(A) - S(B)$, since setting 
variables in $B$ to $r_G$ increases $S(B)$ and decreases the difference.

Thus, to approximate MED-ERR, we adopt a brute force approach,
trying all combinations of up to $p = 5$ bound variables in $A$ down to
depth~30.
We set each combination to the value $r_G$, compute the value of ERR,
and take the maximum across the combinations.
This brute force approximation requires a dozen milliseconds or so on a
typical desktop machine.

\begin{figure*}[p]
\centering
\includegraphics[height=4.5cm]{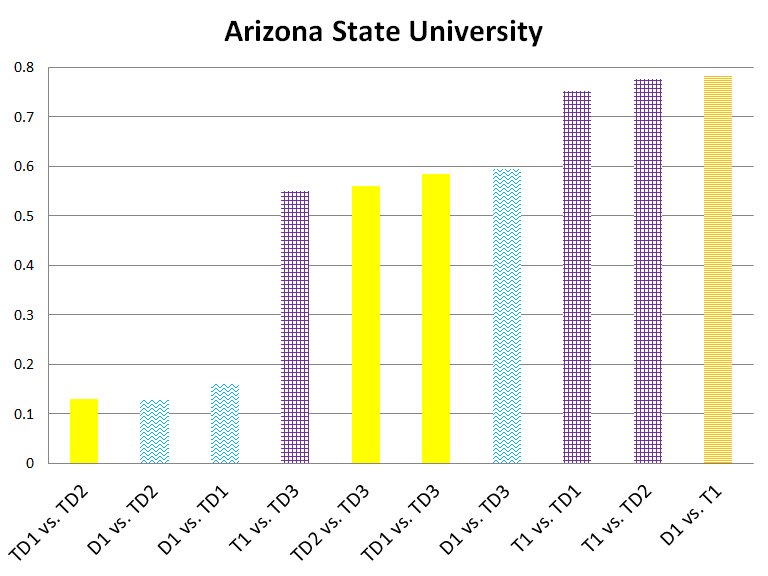}
\hspace*{2.0cm}
\includegraphics[height=4.5cm]{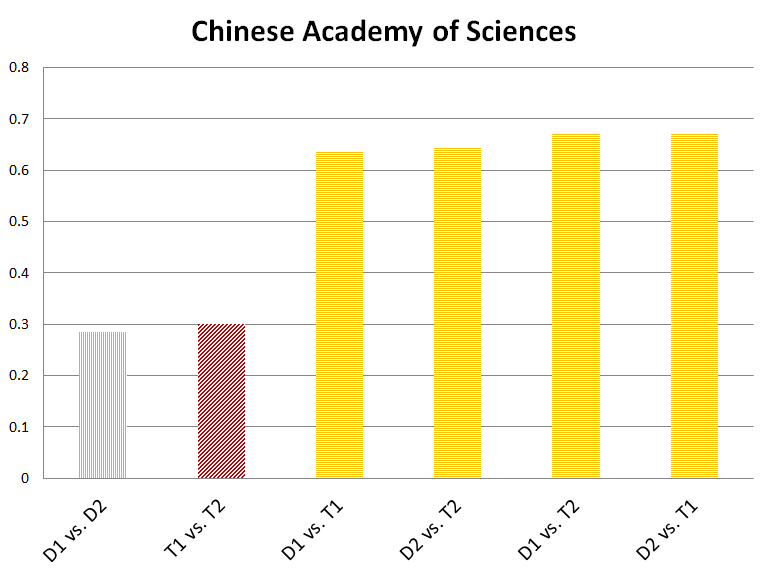}
\vspace*{0.5cm}\\
\includegraphics[height=4.5cm]{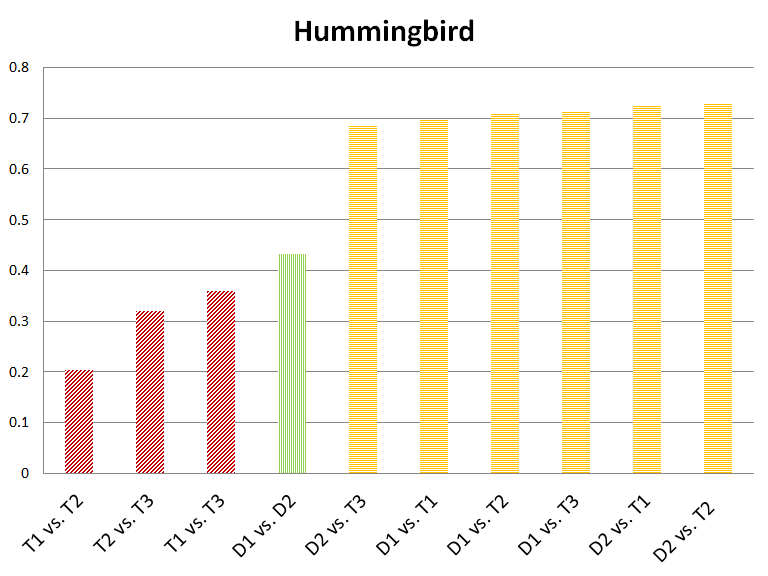}
\hspace*{2.0cm}
\includegraphics[height=4.5cm]{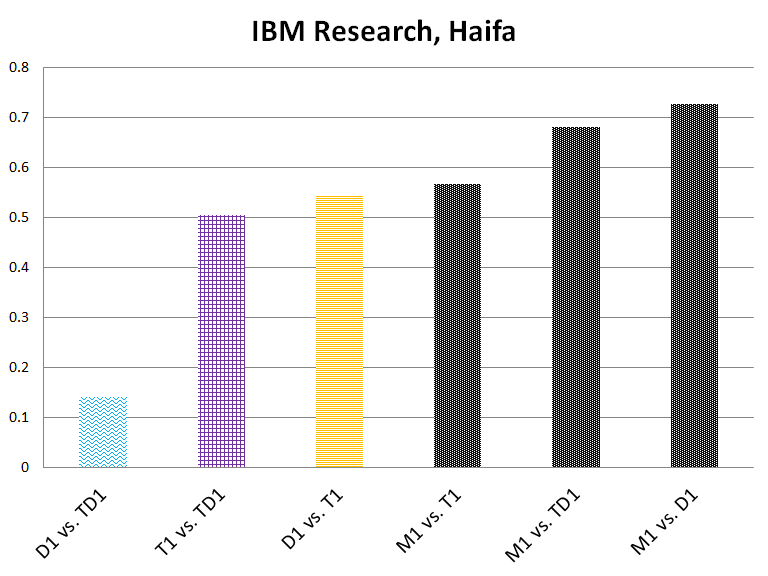}
\vspace*{0.5cm}\\
\includegraphics[height=4.5cm]{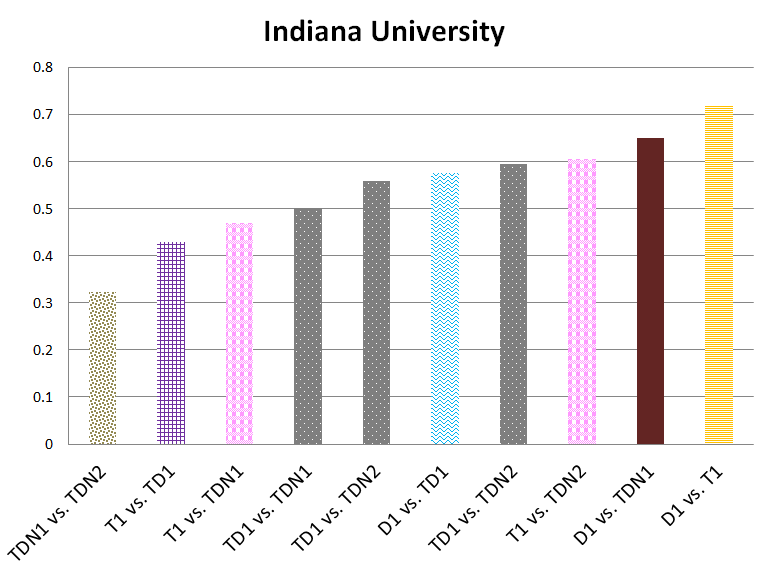}
\hspace*{2.0cm}
\includegraphics[height=4.5cm]{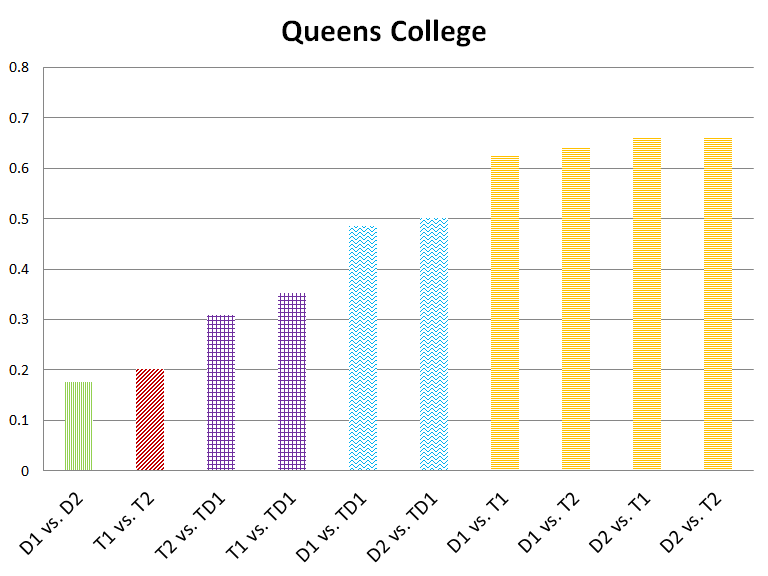}
\vspace*{0.5cm}\\
\includegraphics[height=4.5cm]{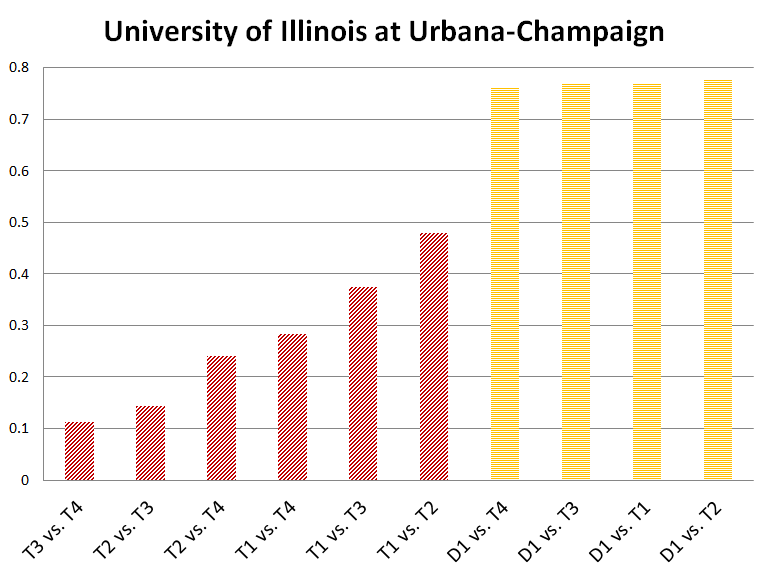}
\hspace*{2.0cm}
\includegraphics[height=4.5cm]{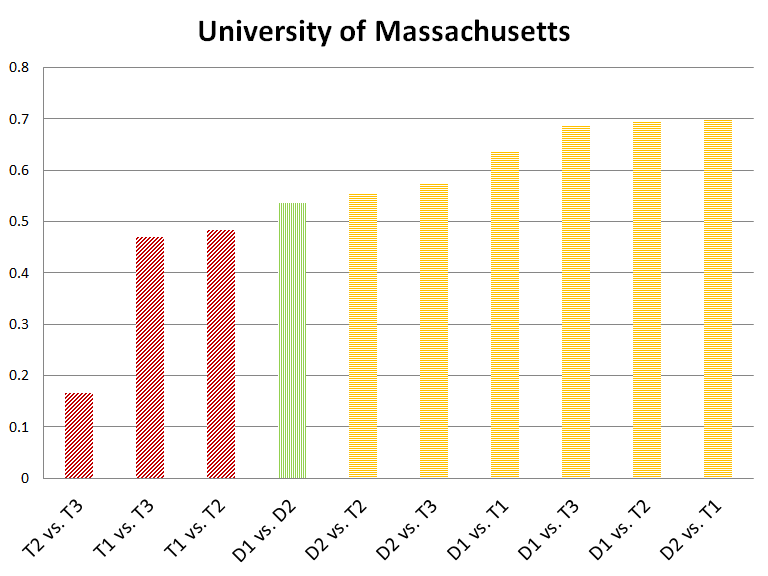}
\\
\caption{
  Intra-group MED-nDCG@20 values for selected TREC 2005 Robust Retrieval Track
  participants.
}
\label{Fwithin}
\end{figure*}

\section{Validation}
\label{SValidation}

To validate MED, we employ a set of experimental runs submitted to the TREC
2005 Robust Retrieval Track~\cite{robust05}.
As our primary goal, we hope to demonstrate that MED reflects meaningful
differences between retrieval runs.
The TREC Robust Track provides a particularly appropriate dataset for this
purpose because its retrieval topics were chosen for their anticipated
difficulty, with many standard retrieval algorithms performing poorly on them.
As a result, track participants applied an unusually wide variety of
retrieval methods, including some entirely novel methods, particularly in
the area of query expansion.

A total of 17 groups participated in the track submitting a total of 74 runs.
We start by examining differences between runs submitted by the same group.
If we assume that each group used a core retrieval approach across all its runs,
we hope to interpret MED values in terms of meaningful changes between these
runs.
Among the groups who submitted multiple runs, we selected the eight groups
with the best overall performance.
While we now focus our attention on these eight groups, we note that results
for the excluded groups exhibit consistent behavior,
with the general observations we make below applying to these groups as well.

\begin{figure*}[tp]
\centering
\subfigure[No predetermined variables] {
  \includegraphics[angle=-90,width=8.5cm]{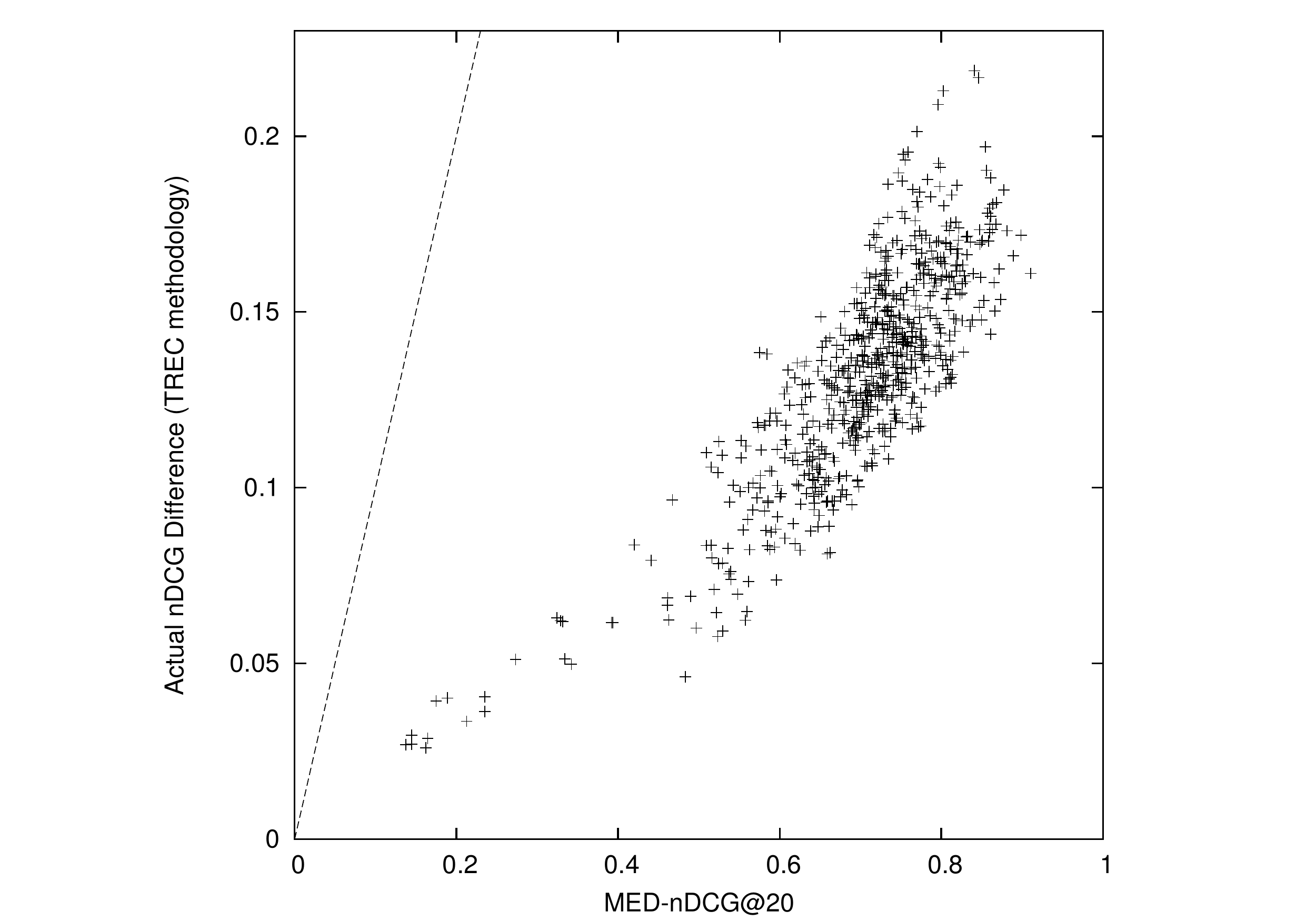}
  \label{Factual00}
}
\subfigure[25\% of available TREC qrels] {
  \includegraphics[angle=-90,width=8.5cm]{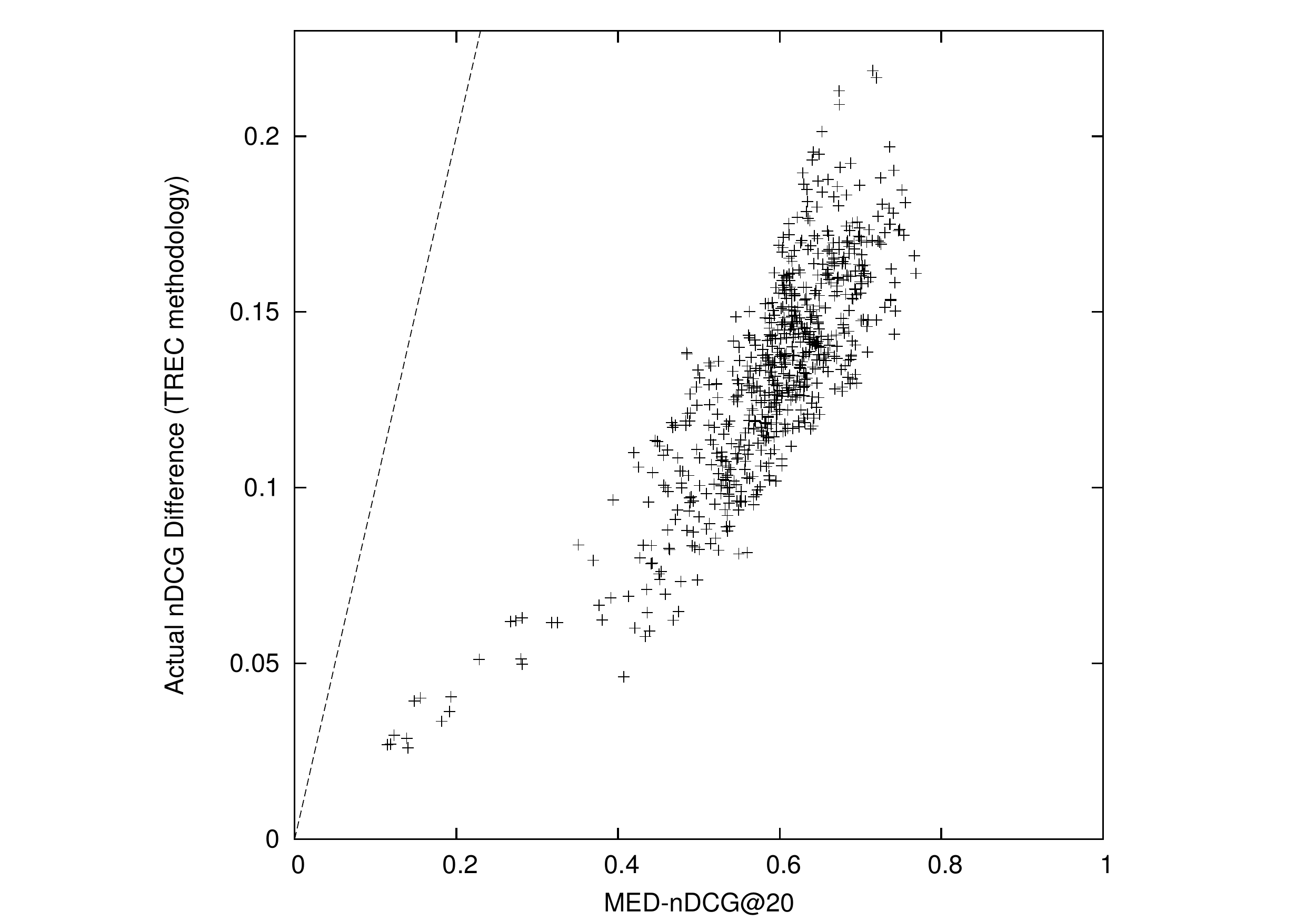}
  \label{Factual25}
}
\subfigure[75\% of available TREC qrels] {
  \includegraphics[angle=-90,width=8.5cm]{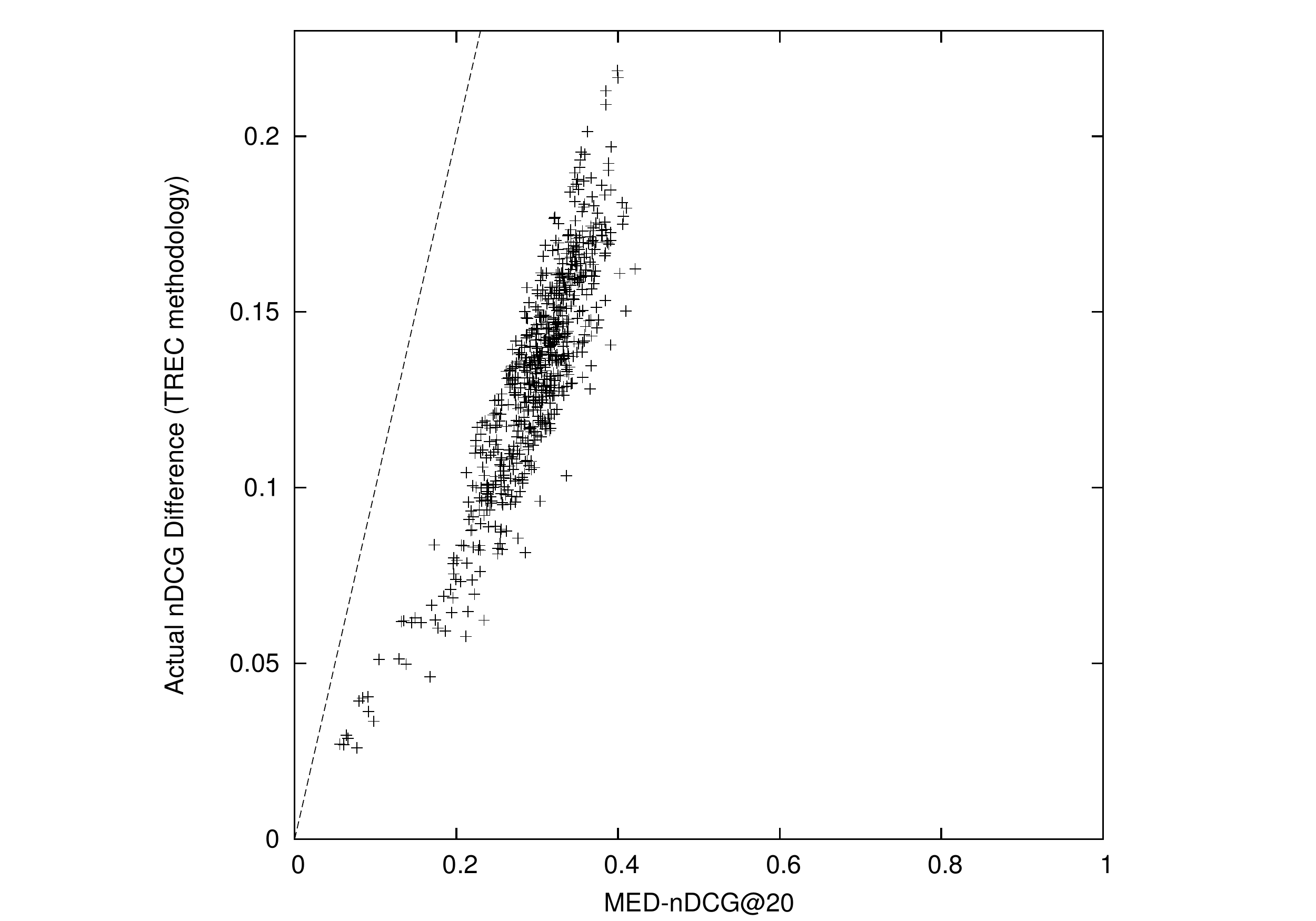}
  \label{Factual75}
}
\subfigure[100\% of available TREC qrels] {
  \includegraphics[angle=-90,width=8.5cm]{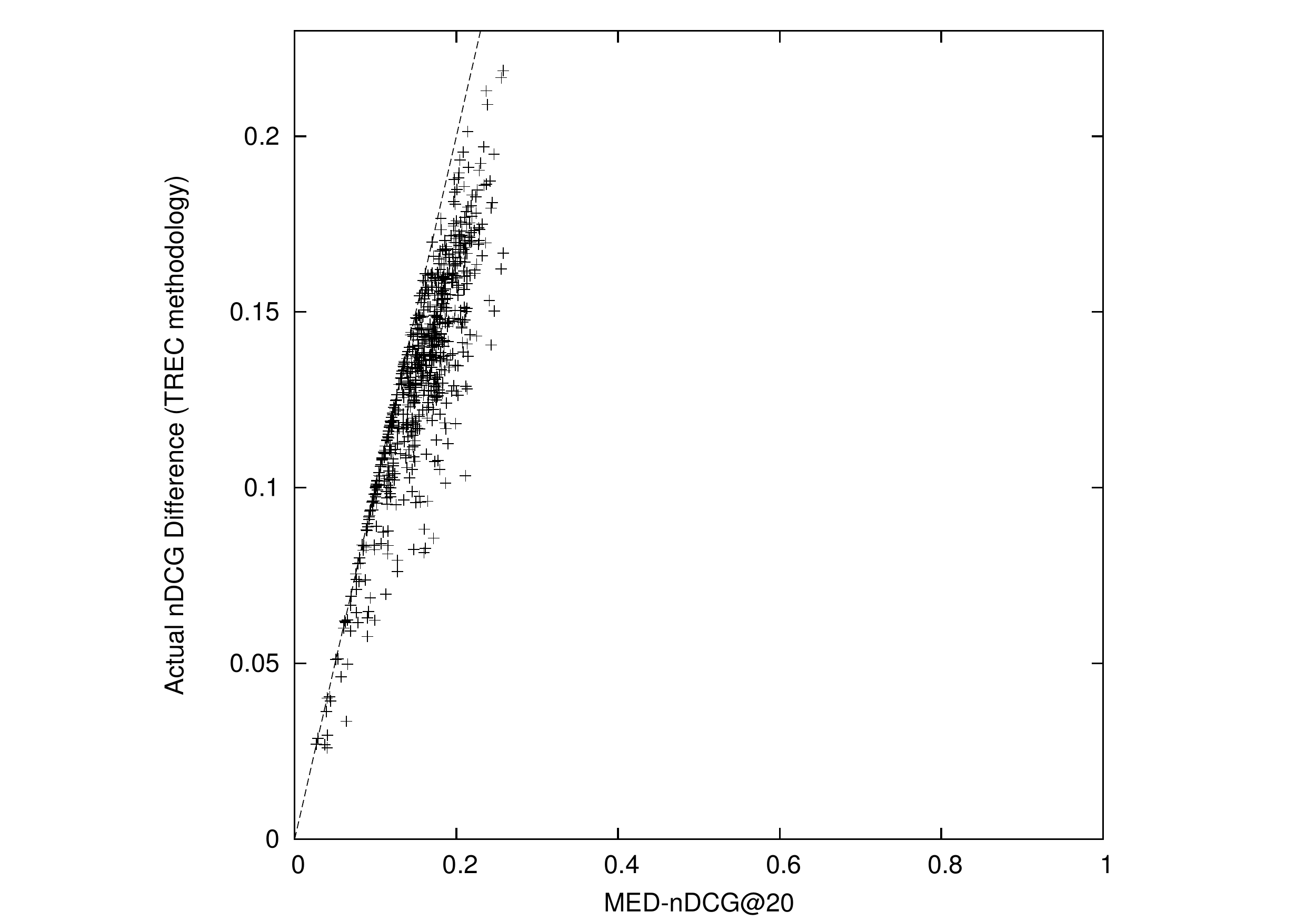}
  \label{Factual100}
}
\caption{
  Actual nDCG differences vs.\ MED-nDCG@20 across all pairs of runs from
  the TREC 2005 Robust Retrieval Track.
}
\label{Factual}
\end{figure*}

We computed MED-nDCG between all pairs of runs within each group, averaging
across the 50 topics used in the track.
Figure~\ref{Fwithin} shows the results, with one chart for each group,
appearing in alphabetical order.
Within each chart, pairs are ordered by increasing MED distance.
Each bar corresponds to a single pair of runs, with the value of MED-nDCG
given on the y-axis.

Each run is identified by a code indicating the type of information used to
automatically formulate the query.
Like many TREC tasks, retrieval topics for the Robust Track provide multiple
expressions of the associated information need, with each topic including:
\begin{itemize}
  \item
    a title field (T), providing one to three keywords expressing the
    information need,
  \item
    a description field (D), providing a single sentence expressing the
    information need,
    and
  \item
    a narrative (N), providing a longer expression of the information need.
\end{itemize}
Run codes indicate the set of fields used for query formulation.
For example, the codes TD1 and TD2 in the Arizona State University chart both
indicate runs where the title and description were used for query formulation.
Within each set of codes, numbers are assigned arbitrarily.
In addition, IBM submitted a single manual run (M1) in which the query
formulation process included human assistance.

We indicate the types of queries in each pair using unique colors and shading.
For example, title vs.\ title runs are colored in strawberry,
description vs.\ description runs are colored in lime,
and title vs.\ description runs are colored in tangerine.
The charts for the Chinese Academy of Science, Hummingbird, Illinois, and
Massachusetts contain only pairs of these types.
An interesting trend is apparent in these four charts, with
title vs.\ title pairs and description vs.\ description pairs being noticeably 
closer than title vs.\ description pairs.

Digging deeper, we consider the runs from Arizona State University.
From the chart, we see that TD1, TD2, and D1 are all several times closer to
each other than they are to the other two runs.
These two runs, T1 and TD3, are closer to each other than they are to any of
the other runs.
The group's workshop report provides an explanation~\cite{rcfr05}.
Reading their report, we learn that T1 and TD3 used the same query expansion
model, which differed from the expansion model used by the other runs.
Digging to the other groups reveals similar relationships reflected in MED-nDCG.
For example, as shown in the chart from IBM, manual runs tend to be farther
from automatic runs than automatic runs are from each other.

The computation of MED does not depend on relevance judgments.
However, we would hope that MED distance provides some indication about  
actual effectiveness differences.
Figure~\ref{Factual00} plots (the absolute value of) actual nDCG@20 differences
against MED-nDCG@20 distances across all pairs for the groups listed in
Figure~\ref{Fwithin}.
The plot includes both intra-group pairs and inter-group pairs,
703 pairs in total.
The plot shows a clear correlation, although the MED-nDCG distances have much
higher values.

As relevance judgments become available, creating predetermined variables,
MED-nDCG distances become increasingly correlated with actual nDCG
differences, moving closer and closer to the actual differences.
Figure~\ref{Factual25} plots actual nDCG differences against MED-nDCG
after we set variables to predetermined values from the
using 25\% of available TREC relevance judgments, randomly selected.
Figure~\ref{Factual75} provides an equivalent plot for 75\% of available TREC
judgments,
and Figure~\ref{Factual100} provides an equivalent plot for 100\% of
available TREC judgments.
In this last plot, all of the MED-nDCG distances do not match actual
differences because TREC 2005 Robust Track runs were not fully judged down
to depth 20, and TREC assumes unjudged documents to be non-relevant.

\begin{figure*}[tp]
\centering
\subfigure[MED-RBP vs.\ RBO]{
  \includegraphics[width=5.6cm]{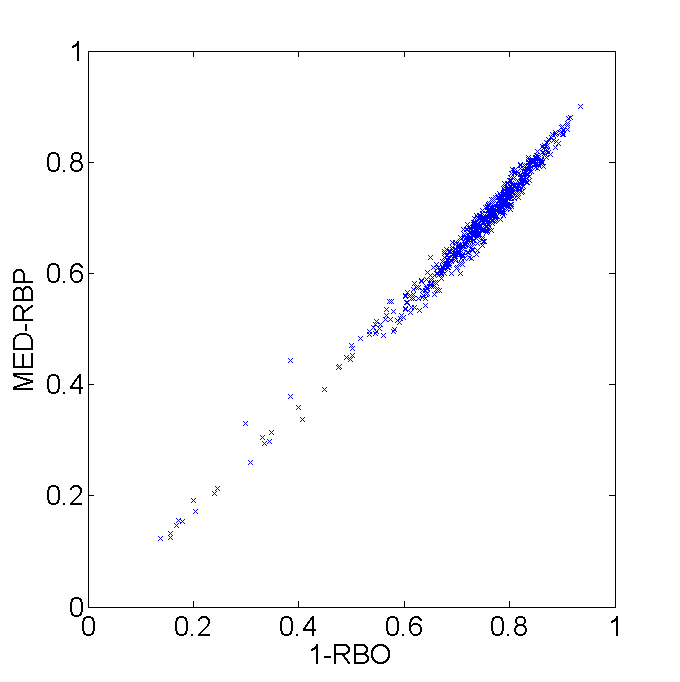}
  \label{Fcomp}
}
\subfigure[MED-RBP vs.\ MED-nDCG@20]{
  \includegraphics[width=5.6cm]{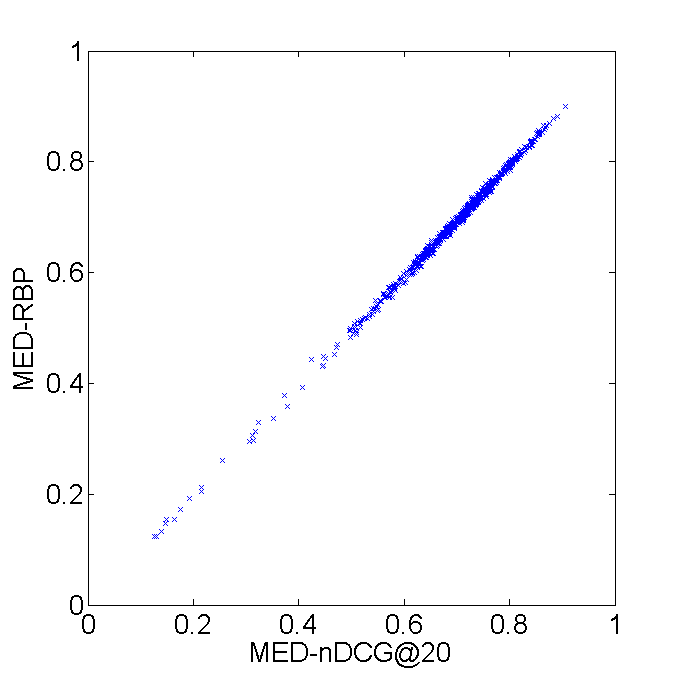}
  \label{FRBP}
}
\subfigure[MED-MAP vs.\ MED-nDCG@20]{
  \includegraphics[width=5.6cm]{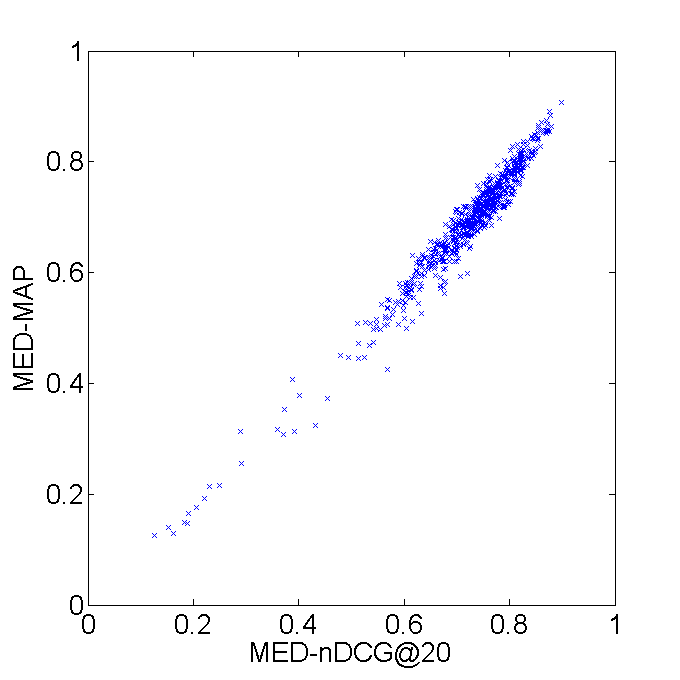}
}
\subfigure[MED-ERR vs.\ MED-nDCG@20]{
  \includegraphics[width=5.6cm]{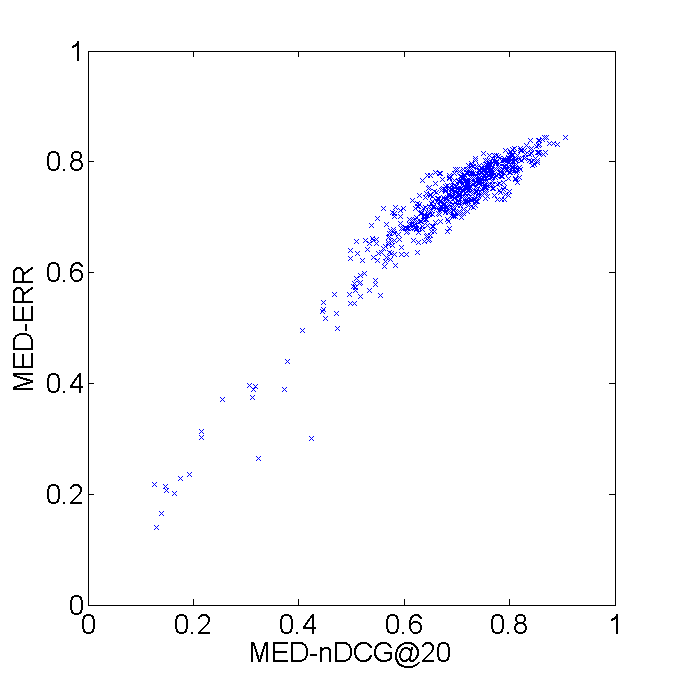}
}
\subfigure[MED-nDCG@5 vs.\ MED-nDCG@20]{
  \includegraphics[width=5.6cm]{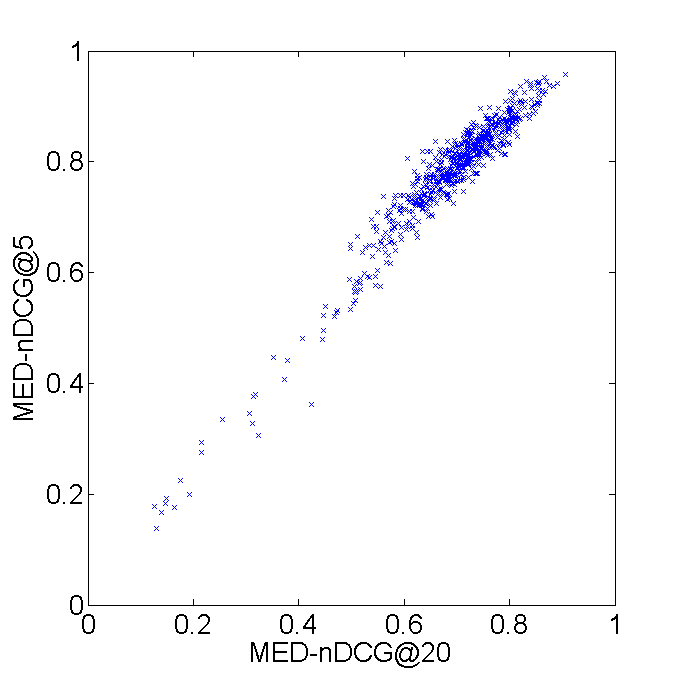}
  \label{F005}
}
\subfigure[MED-nDCG@100 vs.\ MED-nDCG@20]{
  \includegraphics[width=5.6cm]{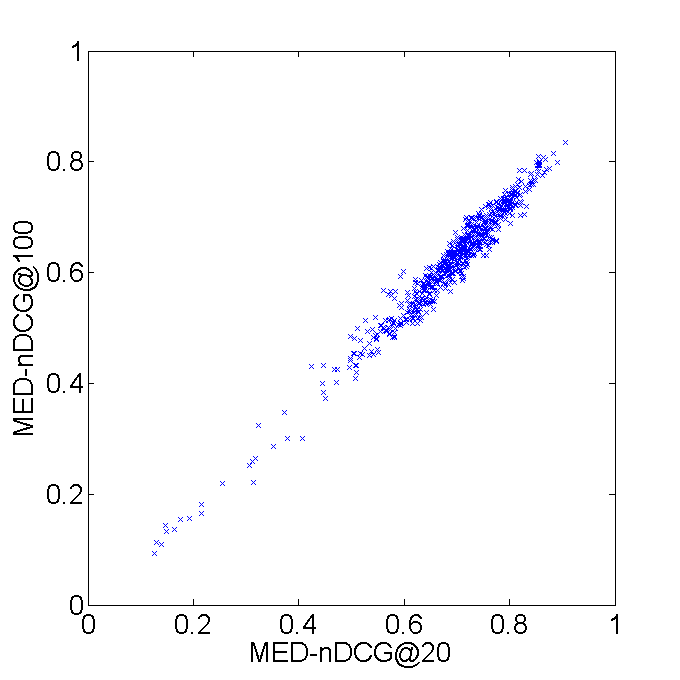}
  \label{F100}
}
\caption{
  Comparisons between RBO, MED-nDCG, MED-RBP, MED-MAP and MED-ERR for all
  runs from the TREC 2005 Robust Retrieval Track.
}
\label{Fcorr}
\end{figure*}

Figure~\ref{Fcorr} provides comparisons between RBO, MED-nDCG, MED-RBP,
MED-MAP and MED-ERR.
In all plots, most of the points fall toward the upper right, indicating
larger differences in the pairs.
The pairs appearing towards the lower left are generally from the same groups,
using the similar retrieval techniques.

Given that MED-RBP and RBO share a common user model, we would expect
a high correlation between them.
Figure~\ref{Fcomp} shows the correlation.
The strength of this correlation suggests that other variants of
MED may similarly reflect the user models underlying those measures.
Interestingly, MED-RBP is even more highly correlated with MED-nDCG@20,
as shown in Figure~\ref{FRBP}.
This correlation is related to the choice of persistence parameter
($\psi = 0.9$).
Lower and higher values for $\psi$ produce weaker correlations.
Figures~\ref{F005} and~\ref{F100} show how values of MED-nDCG change for
different values of $k$ (5, 20, and 100).

\section{Comparison with Prior Work}
\label{SComparison}

As discussed in the introduction, the RBO similarity measure
of Webber et al.~\cite{wmz10} directly inspired our efforts.
As part of that paper, Webber et al.\ provide a substantial review of
related research up to early 2010, when the final version of their paper
was submitted and accepted.
We encourage readers who are interested in a thorough analysis of this prior
work to consult that paper.
In this section, we touch only on the prior work that is most closely
connected with our efforts, including papers appearing since early 2010.

Working independently of Webber et al., and publishing at roughly the same
time, Sun et al.~\cite{slc10} identified a similar set of desiderata for rank
similarity in the context of search.
These desiderata essentially include a requirement for top-weighting,
and an ability to handle indefinite and incomplete results.
In addition, they suggest that measures should be symmetric,
should be computationally efficient,
and should allow meaningful aggregation over multiple queries.
To address these desiderata, Sun et al. defined a similarity measure based on
a weighted version of the Hoeffding distance.
They then applied this measure to visualize differences in search engines
through multidimensional scaling.

MED satisfies the additional desiderata of Sun et al.
Symmetry is straightforward from the definition of MED.
The versions of MED detailed in this paper satisfy the efficiency requirement
to varying degrees, and all are reasonably efficient.
Computation of MED for the dot product measures is essentially instantaneous,
with run times dominated by I/O and data conversion.
For members of the MED family not covered in this paper, computational
efficiency will depend on the details of the associated effectiveness measure.
MED also derives its approach to aggregation from the associated effectiveness
measure, where an arithmetic mean is typical, but other approaches are
possible~\cite{gmap}.

Kumar and Vassilvitskii~\cite{kv10} proposed various extensions to the
Kendall's $\tau$ and Spearman's footrule correlation coefficients intended
to address the requirements of search.
These extensions are intended to handle top weighting, known document relevance,
and the similarity between documents.
They demonstrated that these extensions maintain the Diaconis-Graham inequality,
which guarantees that Kendall's $\tau$ and Spearman's footrule differ
by at most a constant factor.

Like Webber et al.~\cite{wmz10} and Sun et al.~\cite{slc10},
Kumar and Vassilvitskii suggest a list of desiderata for rank similarity
in the context of search.
These desiderata include support for top weighting and the triangle inequality.
While all their requirements are not precisely defined, MED appears to satisfy
them with one interesting exception.
The family members of MED defined in this paper do not provide support for
inter-document similarity.
When comparing two rankings, replacements or swaps of documents with similar
content may not greatly impact the user, and a similarity measure might
reasonably reflect this consideration.
From the perspective of MED, we can trace this requirement back to the
associated effectiveness measures, which also do not appropriately handle
documents with similar or duplicate content.
As researchers begin to consider these issues in the design of effectiveness
measures~\cite{mc12}, solutions will transfer naturally to MED.

Several other researchers have also adapted existing rank similarity measures
to the requirements of search, particularly the Kendall's $\tau$ correlation
coefficient~\cite{yar08,ben09,mel09}.
These efforts primarily address the issue of top weighting, paying less
attention to incomplete and indefinite rankings.
In addition, rather than measuring differences between result lists,
much of this work focuses on the problem of comparing system rankings
for the purpose of validating newly proposed effectiveness measures,
and other evaluation methodologies.
In our work, rather than adapting existing rank similarity measures to the
requirements of search, we adapt existing search effectiveness measures to
the computation of rank similarity, inheriting important properties of these
effectiveness measures and cleanly accommodating known relevance information.  

Rank similarity measures provide a method for comparing search results without
the need for relevance information.
Numerous researchers have examined the related problem of estimating
effectiveness measures and comparing systems using limited relevance
information.
Most notably, Carterette et al.~\cite{ca05,cas06,ben08} define algorithms for
selecting minimal sets of documents for judging, with the aim of ordering a
group of retrieval systems at a given confidence level.
Similar in spirit to MED, and using equivalent methods,
documents are selected to maximize differences in retrieval effectiveness
scores.

Interestingly, in his doctoral thesis Carterette~\cite[pages 156--157]{ben08}
notes the need for rank similarity measures that incorporate top-weighting
and an ability to handle incomplete rankings.
To address these shortcomings of traditional correlation coefficients,
such as Kendall's $\tau$ and Spearman's $\rho$,
he proposes his own rank similarity measure based on differences in
reciprocal rank.
He employs this measure to study properties of his judging algorithms and
TREC runs.

We take his work a step further,
recognizing and demonstrating that maximized effectiveness difference
itself can form the basis for measuring rank similarity.
In addition, we extend the idea of maximized effectiveness difference to
effectiveness measures that did not exist at the time of his thesis,
including ERR and RBP
(and in the next section, measures beyond the ranked list).
Some of our ideas could also be applied back to his work,
including our generalization for the dot-product measures,
our formulation for MAP, and our formulation for ERR.

Apart from Webber et al.~\cite{wmz10} and
Carterette et al.~\cite{ca05,cas06,ben08},
the work closest to ours is the AnchorMap measure proposed by
Buckley~\cite{chris04}.
AnchorMap compares two ranked lists by assuming that the top $k$ documents
from one list are relevant, and then computing MAP for the other list using
this relevance information.
While AnchorMap is not symmetric, and has other limitations~\cite{wmz10},
the idea captures the essence of our proposal.

Several researchers have used rank similarity measures to monitor and compare
commercial search engines~\cite{wmz10,kv10,cm11,slc10}.
Cardoso and Magalh\~{a}es~\cite{cm11} applied RBO to compare the behavior of
two of the most heavily used commercial search engines.
In addition, they applied Jensen-Shannon divergence to measure the similarity
between search results based on the contents of the top documents returned.
They argue that this second approach provides deeper insights into the
differences between the search engines, a view that reflects some concerns
of Kumar and Vassilvitskii~\cite{kv10}.

In other related work,
B\"{u}ttcher et al~\cite{stefan07} trained a classifier to predict the
relevance of unjudged documents.
Jensen et al.~\cite{jbcf07} combined limited manual judgments with automatically
generated pseudo-judgments to evaluate search results in dynamic environments.
Yilmaz et al.~\cite{yka08} developed sampling methods for estimating standard
effectiveness measures.
Sakai and Kando~\cite{sk08} explored the impact of missing relevance judgments
on standard effectiveness measures across four large test collections.

\begin{figure*}[tp]
\centering
\subfigure[Pairs of retrieval results from the same group] {
  \includegraphics[width=8.5cm]{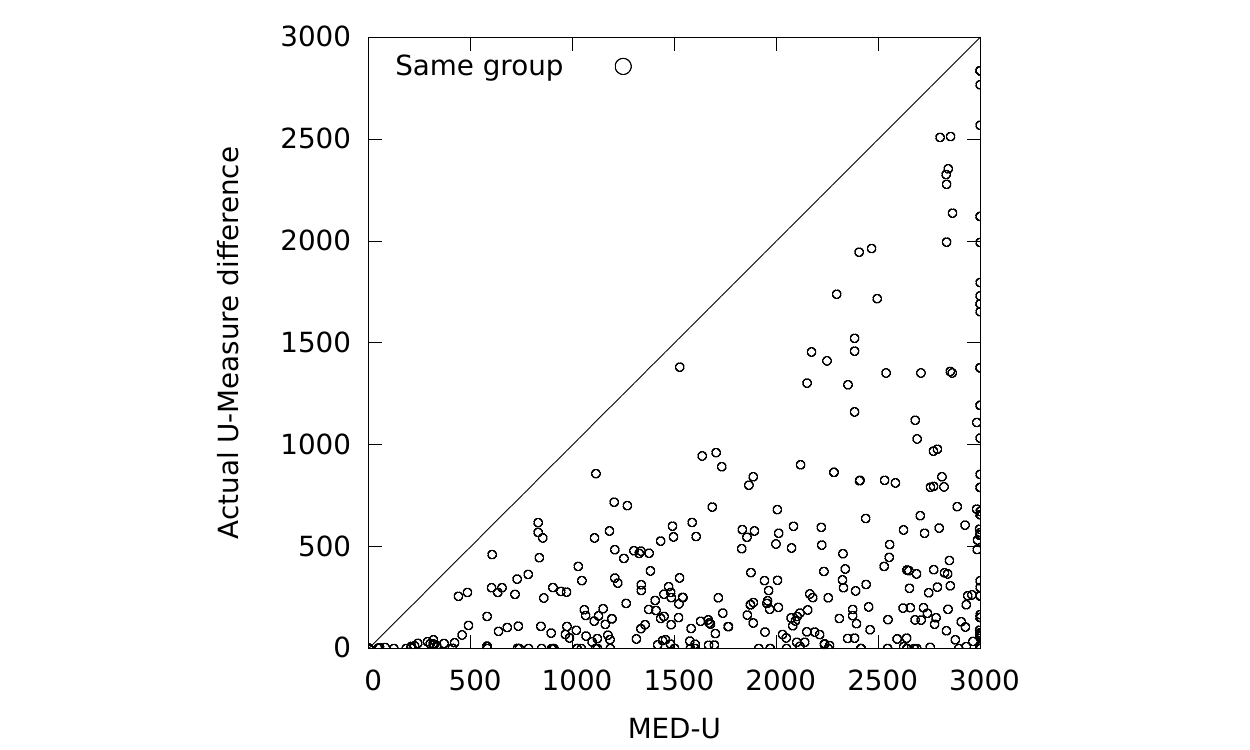}
  \label{FUSame}
}
\subfigure[Pairs of retrieval results from different groups] {
  \includegraphics[width=8.5cm]{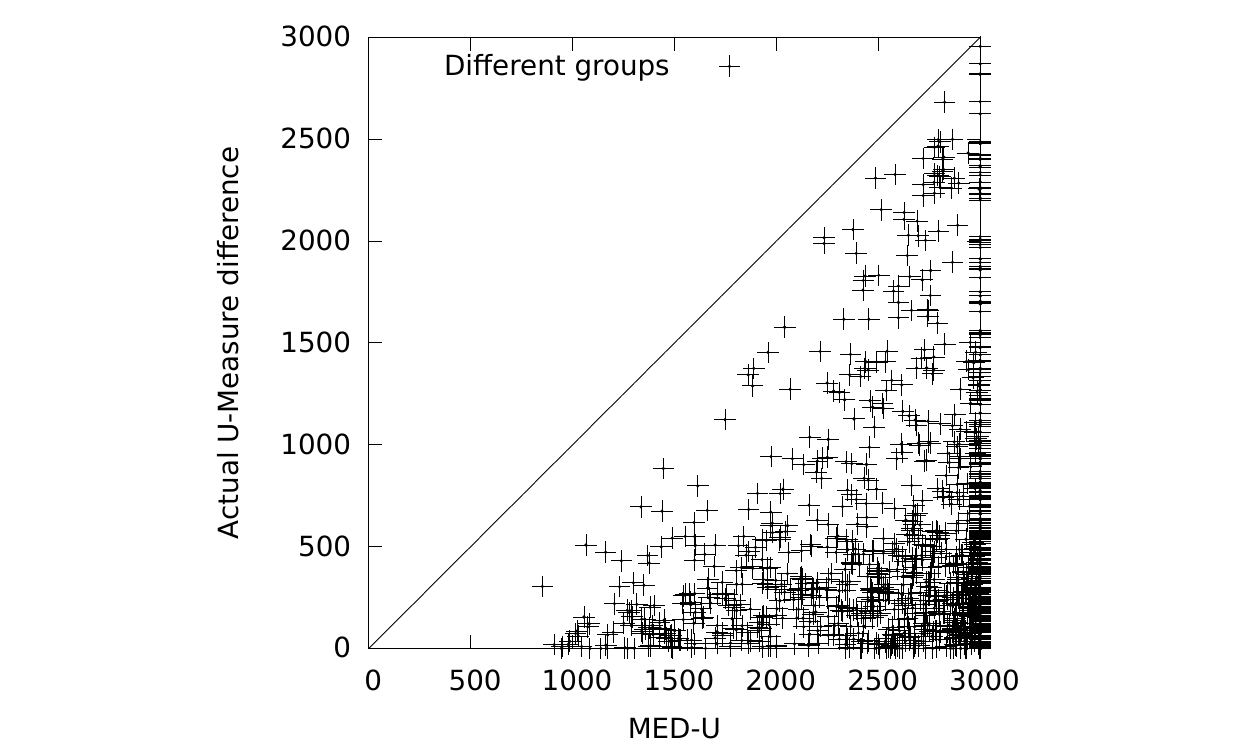}
  \label{FUDiff}
}
\caption{
  Actual U-measure differences vs.\ MED-U@12000 across pairs of
  passage-oriented runs on passage-oriented topics from the
  TREC 2004 HARD Track.
  We plot a 20\% sample, randomly selected, for visual clarity.
}
\label{FU}
\end{figure*}

\section{Beyond the Ranked List}
\label{SBeyond}

MED may be extended to measure distances between search results beyond the
ranked list.
As a simple example, we consider the \emph{U-measure} proposed by
Sakai et al.\cite{sd13}.
This measures assumes that a retrieval result is not expressed as a ranked
list of documents,
but rather as a \emph{trailtext},
a concatenation of all text presented to (or seen by) the user.
Given a trailtext of length $l$, the U-measure is computed as
\begin{equation}
\label{HappyTrailsEqn}
  \frac{1}{{\cal N}} \sum_{i = 1}^{l} c_i d_i.
\end{equation}
Just as Equation~\ref{RBO} sums over documents in the ranked list,
the U-measure sums over character offsets in the trailtext.
The value $c_i$ represents the graded relevance value of the character
at offset $k$,
and the value $d_i$ represents a position-oriented discount.
For their experiments, Sakai and Dou use no normalization (${\cal N} = 1$)
and a linear discount ($d_i = 1 - i/l$)
but clearly the measure can be generalized to other discounts
and normalization values.

The U-measure provides one method for evaluating retrieval systems that
attempt to return sub-document components, such as text passages, to the user.
For example, a system might respond to a query by extracting relevant
passages from books, and other longer documents,
placing them in order for reading by the user.
Essentially this problem was addressed by a task included as a option to the
TREC 2004 HARD Track~\cite{allan04}.

The 2004 HARD Track included 25~topics for which relevance judging was
performed at the sub-document level,
by identifying offsets and lengths of relevant passages within documents.
The corpus for the track comprised more than a half-a-million news articles
from the year 2003, drawn from a variety of sources,
including the Associated Press, the New York Times, and the Washington Post.
Participating systems returned a ranked list of passages for each of the
passage-level topics,
where each passage consisted of a starting byte offset within a document
and a length in bytes.

Since characters within passages cannot be arbitrarily re-ordered,
standard rank correlation coefficients are not appropriate to compute
similarities between these passage-oriented results.
However, we may apply apply MED for the U-measure (MED-U) for this purpose.
Although we have switched from documents to characters,
the properties of MED from Section~\ref{SNotation} continue to hold;
nothing in that section requires variables to represent documents.

The U-measure may be viewed as a type of dot-product measure,
as discussed in Section~\ref{SDotMED}.
We apply the methods of that section to compute MED-U over passage-oriented
retrieval runs submitted to the track.
We compute the measure down to depth $l = 12,000$,
consistent with HARD Track practice,
and use binary values for $c_i$, i.e, a character is either relevant or not.
Results are shown in Figure~\ref{FU}.

Figure~\ref{FU} plots actual U-measure differences vs.\ U-MED across pairs
of passage-oriented topics taken from passage-oriented runs.
For visual clarity, we plot a 20\% sample, selected randomly.
Figure~\ref{FUSame} plots intra-group pairs;
Figure~\ref{FUDiff} plots inter-group pairs.
In contrast to Figure~\ref{Factual00},
retrieval results tend to be very different from one another,
although runs from the same group tend to be closer than runs from different
groups.
Many pairs from different groups are completely different
(with the maximum possible MED-U@12000 value of 2999.75).
These large differences may reflect the unusual nature of the task,
as well as unfamiliarity with passage retrieval.
If passage retrieval was a better understood task,
we might much expect closer results,
particularly for intra-group runs.

In other work,
Smucker and Clarke~\cite{mc12,mc12b,cs14} present a general effectiveness
measure, called \emph{time-biased gain}, which may be applied to measure search
results in many forms.
For example, retrieval results might be specified as summaries, snippets, or
video clips.
As the user interacts with the results, the total benefit to the user is
expressed as a function $G(t)$, the \emph{cumulative gain} at time $t \geq 0$.
This gain is realized as relevant material is encountered by the user,
perhaps by reading relevant text or viewing relevant video.
For a retrieval result $X$, with associated cumulative gain function $G_X(t)$,
time-biased gain is
\begin{equation}
  S(C) \ =\  \int_{0}^{\infty} \frac{dG_X}{dt} D(t) dt.
\end{equation}
$D(t)$ is a \emph{decay} function, typically exponential, indicating the
probability that the user interacts with the material until time $t$.
Given two retrieval results, $A$ and $B$, we treat the material shared by
these results as \emph{bound material}, material appearing in only one
result as \emph{free material}, and material with known relevance as
\emph{predetermined material}.
To compute \mbox{MED}(A,B), we must assign relevance to bound and free
material to maximize
\begin{equation}
  S(A) - S(B) = \int_{0}^{\infty} \frac{d(G_A - G_B)}{dt} D(t) dt\;.
\end{equation}
MED-RBP, MED-nDCG, MED-ERR, and even MED-MAP are all specific
examples of this general equation, under appropriate list-oriented
definitions for gain and decay~\cite{mc12,cs14}.
We leave to future work the application of this equation to measure
search-result distances beyond the ranked list.

\section{Concluding Discussion}
\label{SConclusion}

Rank similarity measures cannot replace traditional measures effectiveness
for determining the absolute performance of search engines.
However, they are more easily applied over larger numbers of queries,
without the need for relevance judgements, providing an additional tool for
assessing the scope of a search engine change.
The MED family of rank similarity measures, presented in this paper,
satisfies various desiderata suggested in prior work for rank similarity
measures in the context of search.
Unlike the rank similarity measures presented in this prior work,
the MED family allows us to translate our understanding and assumptions
about user behavior from an existing effectiveness measure to create a rank
similarity measure.
In addition, MED cleanly accommodates partial relevance judgments,
when this information is available.
If complete relevance information is available, MED reduces to a simple
difference between effectiveness values.
Software to compute MED for various effectiveness measures is available at
{\tt\url{plg.uwaterloo.ca/~claclark/med}}.

Differences between retrieval results might also be studied under
conditions other than the ``worst case'' assumption of MED.
For example, we might assume that an unjudged document is equally likely
to be relevant or non-relevant.
Under this assumption, we can compute a distribution of differences,
perhaps treating the mean difference as a similarity measure,
although it may not be a metric in the mathematical sense.
The probability of relevance for an unjudged document could also be
conditioned on rank, or computed by a classifier, with the goal of
estimating actual differences.
Some of these ideas have been partially explored in related work,
as discussed.
We leave additional exploration as future work.

Some newer proposals for effectiveness measures incorporate more complex user
models, creating interesting and challenging optimization problems,
which we hope to examine in the future.
Several research groups have suggested measures that reward novelty and
diversity in search results~\cite{ss11,cmzg09,ccsa11}.
Computing MED for these effectiveness measures may require the explicit
assignment of query interpretations to documents in order to maximize
effectiveness difference.
Other proposals use time as the primary indicator of user effort,
creating measures that reflect the impact of snippets and other user
interface features~\cite{bkj12,mc12}.
Some of these proposals employed simulation as a method for determining
effectiveness~\cite{bkj12,mc12b,cs14},
and computing MED for these effectiveness measures may also require extensive
simulation.

\section*{Acknowledgements}

We thank Ben Carterette, Yuying Li, Mark Smucker, and William Webber
for helpful feedback at various stages of this research.
Support for this work was provided by Google,
the GRAND Network of Centres of Excellence,
and the Natural Sciences and Engineering Research Council of Canada.

\bibliographystyle{IEEEtran}
\bibliography{IEEEabrv,ranksim}

\begin{thebibliography}{10}
\providecommand{\url}[1]{#1}
\csname url@samestyle\endcsname
\providecommand{\newblock}{\relax}
\providecommand{\bibinfo}[2]{#2}
\providecommand{\BIBentrySTDinterwordspacing}{\spaceskip=0pt\relax}
\providecommand{\BIBentryALTinterwordstretchfactor}{4}
\providecommand{\BIBentryALTinterwordspacing}{\spaceskip=\fontdimen2\font plus
\BIBentryALTinterwordstretchfactor\fontdimen3\font minus
  \fontdimen4\font\relax}
\providecommand{\BIBforeignlanguage}[2]{{%
\expandafter\ifx\csname l@#1\endcsname\relax
\typeout{** WARNING: IEEEtran.bst: No hyphenation pattern has been}%
\typeout{** loaded for the language `#1'. Using the pattern for}%
\typeout{** the default language instead.}%
\else
\language=\csname l@#1\endcsname
\fi
#2}}
\providecommand{\BIBdecl}{\relax}
\BIBdecl

\bibitem{kk02}
K.~J\"{a}rvelin and J.~Kek\"{a}l\"{a}inen, ``Cumulated gain-based evaluation of
  {IR} techniques,'' \emph{ACM Transactions on Information Systems}, vol.~20,
  no.~4, pp. 422--446, 2002.

\bibitem{cmzg09}
O.~Chapelle, D.~Metlzer, Y.~Zhang, and P.~Grinspan, ``Expected reciprocal rank
  for graded relevance,'' in \emph{18th ACM Conference on Information and
  Knowledge Management}, Hong Kong, 2009, pp. 621--630.

\bibitem{najork07}
M.~A. Najork, ``Comparing the effectiveness of {HITS} and {SALSA},'' in
  \emph{16th ACM Conference on Information and Knowledge Management}, Lisbon,
  Portugal, 2007, pp. 157--164.

\bibitem{cz09}
O.~Chapelle and Y.~Zhang, ``A dynamic {Bayesian} network click model for web
  search ranking,'' in \emph{18th International World wide Web Conference},
  Madrid, 2009, pp. 1--10.

\bibitem{dl10}
G.~Dupret and C.~Liao, ``A model to estimate intrinsic document relevance from
  the clickthrough logs of a web search engine,'' in \emph{3rd ACM
  International Conference on Web Search and Data Mining}, New York, 2010, pp.
  181--190.

\bibitem{ehvs12}
C.~Eickhoff, C.~G. Harris, A.~P. de~Vries, and P.~Srinivasan, ``Quality through
  flow and immersion: {G}amifying crowdsourced relevance assessments,'' in
  \emph{35th International ACM SIGIR Conference on Research and Development in
  Information Retrieval}, Portland, Oregon, 2012, pp. 871--880.

\bibitem{hzcwy11}
B.~Hu, Y.~Zhang, W.~Chen, G.~Wang, and Q.~Yang, ``Characterizing search intent
  diversity into click models,'' in \emph{20th International World Wide Web
  Conference}, Hyderabad, 2011, pp. 17--26.

\bibitem{jgphg05}
T.~Joachims, L.~Granka, B.~Pan, H.~Hembrooke, and G.~Gay, ``Accurately
  interpreting clickthrough data as implicit feedback,'' in \emph{28th Annual
  International ACM SIGIR Conference on Research and Development in Information
  Retrieval}, Salvador, Brazil, 2005, pp. 154--161.

\bibitem{sj10}
M.~D. Smucker and C.~P. Jethani, ``Human performance and retrieval precision
  revisited,'' in \emph{33rd International ACM SIGIR Conference on Research and
  Development in Information Retrieval}, Geneva, 2010, pp. 595--602.

\bibitem{chris04}
C.~Buckley, ``Topic prediction based on comparative retrieval rankings,'' in
  \emph{27th Annual International ACM SIGIR Conference on Research and
  Development in information retrieval}, Sheffield, 2004, pp. 506--507.

\bibitem{slc10}
M.~Sun, G.~Lebanon, and K.~Collins-Thompson, ``Visualizing differences in web
  search algorithms using the expected weighted {Hoeffding} distance,'' in
  \emph{19th International World Wide Web Conference}, Raleigh, North Carolina,
  2010, pp. 931--940.

\bibitem{mel09}
M.~Melucci, ``Weighted rank correlation in information retrieval evaluation,''
  in \emph{5th Asia Information Retrieval Symposium on Information Retrieval
  Technology}, Sapporo, Japan, 2009, pp. 75--86.

\bibitem{kv10}
R.~Kumar and S.~Vassilvitskii, ``Generalized distances between rankings,'' in
  \emph{19th International World Wide Web Conference}, Raleigh, North Carolina,
  2010, pp. 571--580.

\bibitem{wmz10}
W.~Webber, A.~Moffat, and J.~Zobel, ``A similarity measure for indefinite
  rankings,'' \emph{ACM Transactions on Information Systems}, vol.~28, no.~4,
  pp. 20:1--20:38, November 2010.

\bibitem{ben09}
B.~Carterette, ``On rank correlation and the distance between rankings,'' in
  \emph{32nd International ACM SIGIR Conference on Research and Development in
  Information Retrieval}, Boston, 2009, pp. 436--443.

\bibitem{yar08}
E.~Yilmaz, J.~A. Aslam, and S.~Robertson, ``A new rank correlation coefficient
  for information retrieval,'' in \emph{31st Annual International ACM SIGIR
  Conference on Research and Development in information retrieval}, Singapore,
  2008, pp. 587--594.

\bibitem{jr06}
B.~J. Jansen and M.~Resnick, ``An examination of searcher's perceptions of
  nonsponsored and sponsored links during ecommerce web searching,''
  \emph{Journal of the American Society for Information Science and
  Technology}, vol.~57, no.~14, pp. 1949--1961, 2006.

\bibitem{mz08}
A.~Moffat and J.~Zobel, ``Rank-biased precision for measurement of retrieval
  effectiveness,'' \emph{ACM Transactions on Information Systems}, vol.~27,
  no.~1, pp. 2:1--2:27, December 2008.

\bibitem{ca05}
B.~Carterette and J.~Allan, ``Incremental test collections,'' in \emph{14th ACM
  International Conference on Information and Knowledge Management}, Bremen,
  Germany, 2005, pp. 680--687.

\bibitem{cas06}
B.~Carterette, J.~Allan, and R.~Sitaraman, ``Minimal test collections for
  retrieval evaluation,'' in \emph{29th Annual International ACM SIGIR
  Conference on Research and Development in Information Retrieval}, Seattle,
  2006, pp. 268--275.

\bibitem{ben08}
B.~Carterette, ``Low-cost and robust evaluation of information retrieval
  systems,'' Ph.D. dissertation, University of Massachusetts Amherst, 2008.

\bibitem{cztr08}
N.~Craswell, O.~Zoeter, M.~Taylor, and B.~Ramsey, ``An experimental comparison
  of click position-bias models,'' in \emph{1st International Conference on Web
  Search and Data Mining}, Palo Alto, 2008, pp. 87--94.

\bibitem{ccsa11}
C.~L. Clarke, N.~Craswell, I.~Soboroff, and A.~Ashkan, ``A comparative analysis
  of cascade measures for novelty and diversity,'' in \emph{4th ACM
  international conference on web search and data mining}, Hong Kong, 2011, pp.
  75--84.

\bibitem{mst12}
A.~Moffat, F.~Scholer, and P.~Thomas, ``Models and metrics: {IR} evaluation as
  a user process,'' in \emph{17th Australasian Document Computing Symposium},
  Dunedin, New Zealand, 2012, pp. 47--54.

\bibitem{rrw09}
F.~Rendl, G.~Rinaldi, and A.~Wiegele, ``Solving max-cut to optimality by
  intersecting semidefinite and polyhedral relaxations,'' \emph{Mathematical
  Programming}, vol. 121, no.~2, pp. 307--335, July 2009.

\bibitem{bea98}
J.~E. Beasley, ``Heuristic algorithms for the unconstrained binary quadratic
  programming problem,'' The Management School, Imperial College, London, Tech.
  Rep., December 1998.

\bibitem{robust05}
E.~M. Voorhees, ``Overview of the {TREC} 2005 robust retrieval track,'' in
  \emph{14th Text REtrieval Conference}, Gaithersburg, Maryland, 2005.

\bibitem{rcfr05}
D.~Roussinov, M.~Chau, E.~Filatova, and J.~A. Robles-Flores, ``Building on
  redundancy: {Factoid} question answering and the ``other'','' in \emph{14th
  Text REtrieval Conference}, Gaithersburg, Maryland, 2005.

\bibitem{gmap}
S.~Robertson, ``On {GMAP}: and other transformations,'' in \emph{15th ACM
  International Conference on Information and Knowledge management}, 2006, pp.
  78--83.

\bibitem{mc12}
M.~D. Smucker and C.~L. Clarke, ``Time-based calibration of effectiveness
  measures,'' in \emph{35th International ACM SIGIR Conference on Research and
  Development in information retrieval}, Portland, Oregon, 2012, pp. 95--104.

\bibitem{cm11}
B.~Cardoso and J.~a. Magalh\~{a}es, ``{Google}, {Bing} and a new perspective on
  ranking similarity,'' in \emph{20th ACM International Conference on
  Information and Knowledge Management}, Glasgow, 2011, pp. 1933--1936.

\bibitem{stefan07}
S.~B\"{u}ttcher, C.~L.~A. Clarke, P.~C.~K. Yeung, and I.~Soboroff, ``Reliable
  information retrieval evaluation with incomplete and biased judgements,'' in
  \emph{30th Annual International ACM SIGIR Conference on Research and
  Development in Information Retrieval}, Amsterdam, 2007, pp. 63--70.

\bibitem{jbcf07}
E.~C. Jensen, S.~M. Beitzel, A.~Chowdhury, and O.~Frieder, ``Repeatable
  evaluation of search services in dynamic environments,'' \emph{ACM
  Transactions on Information Systems}, vol.~26, no.~1, November 2007.

\bibitem{yka08}
E.~Yilmaz, E.~Kanoulas, and J.~A. Aslam, ``A simple and efficient sampling
  method for estimating {AP} and {NDCG},'' in \emph{31st Annual International
  ACM SIGIR Conference on Research and Development in Information Retrieval},
  Singapore, 2008, pp. 603--610.

\bibitem{sk08}
T.~Sakai and N.~Kando, ``On information retrieval metrics designed for
  evaluation with incomplete relevance assessments,'' \emph{Information
  Retrieval}, vol.~11, no.~5, pp. 447--470, October 2008.

\bibitem{sd13}
T.~Sakai and Z.~Dou, ``Summaries, ranked retrieval and sessions: {A} unified
  framework for information access evaluation,'' in \emph{36th International
  ACM SIGIR Conference on Research and Development in Information Retrieval},
  Dublin, Ireland, 2013, pp. 473--482.

\bibitem{allan04}
J.~Allan, ``{HARD} track overview in {TREC} 2004: {High} accuracy retrieval
  from documents,'' in \emph{13th Text REtrieval Conference}, Gaithersburg,
  Maryland, 2004.

\bibitem{mc12b}
M.~D. Smucker and C.~L.~A. Clarke, ``Stochastic simulation of time-biased
  gain,'' in \emph{21st ACM International Conference on Information and
  Knowledge Management}, Maui, Hawaii, 2012, pp. 2040--2044.

\bibitem{cs14}
C.~L.~A. Clarke and M.~D. Smucker, ``Time well spent,'' in \emph{Information
  Interaction in Context Conference}, Regensburg, Germany, 2014.

\bibitem{ss11}
T.~Sakai and R.~Song, ``Evaluating diversified search results using per-intent
  graded relevance,'' in \emph{34th International ACM SIGIR Conference on
  Research and Development in Information Retrieval}, Beijing, 2011, pp.
  1043--1052.

\bibitem{bkj12}
F.~Baskaya, H.~Keskustalo, and K.~J\"{a}rvelin, ``Time drives interaction:
  {Simulating} sessions in diverse searching environments,'' in \emph{35th
  International ACM SIGIR Conference on Research and Development in Information
  Retrieval}, Portland, Oregon, 2012, pp. 105--114.

\end{thebibliography}

\end{document}